\documentclass[aps,prd,amsmath,amssymb,preprint,showpacs,superscriptaddress]{revtex4-2}
\usepackage{amsmath}
\usepackage[caption=false]{subfig}
\usepackage{xcolor}
\usepackage{graphicx}
\usepackage{bm}
\usepackage{siunitx,multirow}
\usepackage{makecell}
\usepackage{afterpage}
\usepackage{soul}
\usepackage{dcolumn}

\usepackage[colorlinks=true,linkcolor=blue,allcolors=blue]{hyperref}%
\newcolumntype{M}[1]{>{\centering\arraybackslash}m{#1}}
\newcommand{\aftertwo}[1]{\afterpage{\if@firstcolumn #1
  \else\afterpage{#1}\fi}}

\begin{document}

\title[Article Title]{Evidence of electronic instability driven structural distortion in the nodal line semimetal CoSn$_2$}

\author{Suman Nandi}
\affiliation{Department of Condensed Matter Physics
and Materials Science, Tata Institute of Fundamental
Research, Mumbai 400005, India}

\author{Bishal Maity}
\affiliation{Department of Condensed Matter Physics
and Materials Science, Tata Institute of Fundamental
Research, Mumbai 400005, India}

\author{Shovan Dan}
\affiliation{Department of Condensed Matter Physics
and Materials Science, Tata Institute of Fundamental
Research, Mumbai 400005, India}

\author{Khadiza Ali}
\affiliation{MAX IV Laboratory, Lund University, Lund, SE-221 00, Sweden}

\author{Bikash Patra}
\affiliation{Department of Condensed Matter Physics
and Materials Science, Tata Institute of Fundamental
Research, Mumbai 400005, India}
\affiliation{Department of Physical and Applied Sciences, Indian Institute of Information Technology, Surat, Gujarat
394190, India}

\author{Anshuman Mondal}
\affiliation{ESRF -- European Synchrotron Radiation Facility, 38000 Cedex Grenoble, France}

\author{Gaston Garbarino}
\affiliation{ESRF -- European Synchrotron Radiation Facility, 38000 Cedex Grenoble, France}

\author{Pierre Rodi\`ere}
\affiliation{Institut N\'{e}el CNRS/UGA UPR2940, 25 Rue des Martyrs, 38042 Grenoble, France}

\author{Sitaram Ramakrishnan}
\email{niranj002@gmail.com}
\affiliation{Institut N\'{e}el CNRS/UGA UPR2940, 25 Rue des Martyrs, 38042 Grenoble, France}

\author{Bahadur Singh}
\affiliation{Department of Condensed Matter Physics
and Materials Science, Tata Institute of Fundamental
Research, Mumbai 400005, India}

\author{Arumugam Thamizhavel}
\email{thamiz@tifr.res.in}
\affiliation{Department of Condensed Matter Physics
and Materials Science,
Tata Institute of Fundamental Research,
Mumbai 400005, India}

\date{\today}

\begin{abstract}
Understanding the mechanisms that drive spontaneous rotational symmetry breaking in correlated electron systems is a central challenge in condensed matter physics. Although such symmetry breaking phases have been studied in low-dimensional and strongly correlated materials, its emergence in structurally simpler compounds remains less explored. Here, we investigate non-magnetic CoSn$_2$ that is a centrosymmetric intermetallic compound crystallizing in a tetragonal structure at ambient conditions, and discover an electronically driven symmetry breaking instability. Electrical resistivity reveals a distinct change in the slope below 25 K, deviating from the expected Bloch-Grüneisen behavior. This anomaly is attributed towards a structural change as at 22 K single crystal X-ray diffraction using synchrotron radiation uncovers weak superlattice reflections that leads to a doubling of \textbf{a} and \textbf{c}, resulting in a 4-fold superstructure. The symmetry of the lattice reduces from tetragonal to acentric monoclinic but without any discernible monoclinic distortion down to 10 K. This structural transition is accompanied by a twofold symmetry in angular magnetoresistance, contrasting the fourfold symmetry observed at higher temperatures.  First-principles calculations show no phonon softening but reveal enhanced electronic susceptibility, suggesting an electronic instability. Polarization-dependent ARPES measurements further identify a strong orbital anisotropy dominated by the in-plane Co-$d_{xy}$ states. Collectively, our results point to an electronic instability driven structural distortion in CoSn$_2$, offering a rare platform to study symmetry breaking in a non-magnetic metallic system.
\end{abstract}



\maketitle
\newpage

\section{Introduction}\label{sec1}

Spontaneous symmetry breaking is a fundamental concept in physics, describing situations where a system adopts a symmetry lower than that dictated by its underlying laws or lattice structure. In condensed matter systems, such instabilities often are associated with emergent quantum phenomena like magnetism, superconductivity, charge density waves (CDWs), spin density waves (SDWs), and electronic nematic phases~\cite{blundell2001magnetism, tinkham1975introduction, Gruner1994, Fernandes2014}. In magnetic systems, the continuous spin rotational symmetry is broken as magnetic moments align along a preferred direction, leading to ferromagnetic or antiferromagnetic order~\cite{blundell2001magnetism}. In superconductors, gauge symmetry is spontaneously broken when electrons form Cooper pairs, resulting in a collective quantum state with zero resistance~\cite{tinkham1975introduction}. In layered transition metal dichalcogenides (TMDCs), such as 1T-TiSe$_2$, 2H-TaSe$_2$, and 2H-NbSe$_2$, symmetry breaking transitions manifest due to charge density waves~\cite{PhysRevLett.32.882,PhysRevB.14.4321, doi:10.1143/JPSJ.49.898,PhysRevLett.34.1164}. The microscopic mechanisms are often associated with Fermi surface nesting (FSN) in the case where the geometry of the Fermi surface is simple as in quasi-1D and quasi-2D systems.  In the case of 3D systems FSN may not be realized due to the curvature of the Fermi surface and other mechanisms like strong $q$-dependent electron-phonon interactions are more predominant \cite{PhysRevB.73.205102,PhysRevB.99.174110,PhysRevB.99.161119,ramakrishnan2019a, ramakrishnan2020a, ramakrishnan2023a}. For instance in complex superconducting skutterudite La$_3$Co$_4$Sn$_{13}$, a second-order commensurate CDW forms below 150 K, reducing the crystal symmetry and reconstructing the band structure \cite{welsch2019a}. 
However, a more subtle and enigmatic form of symmetry breaking is realized in electronic nematic phases, where rotational symmetry is broken while preserving translational invariance. The electronic structure develops a preferred direction, thereby leading to anisotropic responses in transport, spectroscopy, or other physical observables, despite the underlying lattice remaining nearly symmetric. This has been extensively explored in iron based superconductors such as FeSe~\cite{PhysRevB.91.201105,PhysRevB.91.155106} and BaFe$_2$(As$_{1-x}$P$_x$)$_2$~\cite{Kasahara2012}, particularly in their high-temperature phases. In URu$_2$Si$_2$~\cite{doi:10.1126/science.1197358,doi:10.1126/science.1259729,Riggs2015}, a hidden order phase below 17.5 K breaks four-fold symmetry without large magnetic moments, suggesting an electronic nematic origin. Similarly, CeRhIn$_5$ exhibits pronounced in plane resistivity anisotropy under strong magnetic fields near its antiferromagnetic quantum critical point, indicative of an electronic nematic phase~\cite{Helm2020}. 

A particularly rich platform for symmetry breaking phenomena has emerged in the family of Kagome materials, which feature a lattice geometry that hosts Dirac like dispersions, flat bands, and van Hove singularities~\cite{Yin2022,Wang2025}. Prominent examples include AV$_3$Sb$_5$ (A = K, Rb, Cs), which exhibit charge density wave transitions, time-reversal symmetry breaking, and nematicity -various intertwined orders that emerge despite the absence of magnetic ordering~\cite{10.1093/nsr/nwac199,hu2023}. For example, in CsV$_3$Sb$_5$, the CDW transition at 94 K is accompanied by a spontaneous breakdown of the sixfold rotational symmetry of the Kagome lattice, leading to a $C_2$ symmetric ground state~\cite{Asaba2024,Xiang2021}. However, the precise temperature at which nematicity onsets remains controversial. In contrast, CsTi$_3$Bi$_5$ exhibits no CDW but still develops a nematic ground state and superconductivity at low temperatures~\cite{Yang2024,PhysRevB.111.045135}. Transitions into a nematic phase are often second-order and may involve only weak structural distortions that are difficult to detect with standard probes such as specific heat. In these systems, evidence for nematicity has largely come from anisotropies observed in transport measurements, polarized angle-resolved photoemission spectroscopy (ARPES), scanning tunneling microscopy (STM), and elastoresistivity, all of which point to an electronically driven instability rather than a purely structural origin.

Despite these advances, the emergence of such symmetry broken states in structurally simple, three dimensional intermetallic compounds remains far less understood. These systems, typically characterized by high lattice symmetry, metallic conductivity, and relatively weak electronic correlations, are often overlooked because symmetry lowered ground states produce changes that are undetectable without highly sensitive probes. In this context we present our work on a less known inter-metallic compound CoSn$_2$ that is primarily applicable for battery storage \cite{park2019a}. Our investigations lead to the uncovering of a symmetry broken state in a structurally simple, three dimensional material which is unprecedented.
 Our findings from the X-ray diffraction in conjunction with angle dependent magnetotransport and polarization resolved ARPES, unravels exotic physics that is associated nematic electronic behavior below 25~K.

\section{Results and discussion}\label{sec2}
\subsection{Physical properties and X-ray diffraction}
\begin{figure*}[!ht]
		\centering
		\includegraphics[scale=0.55]{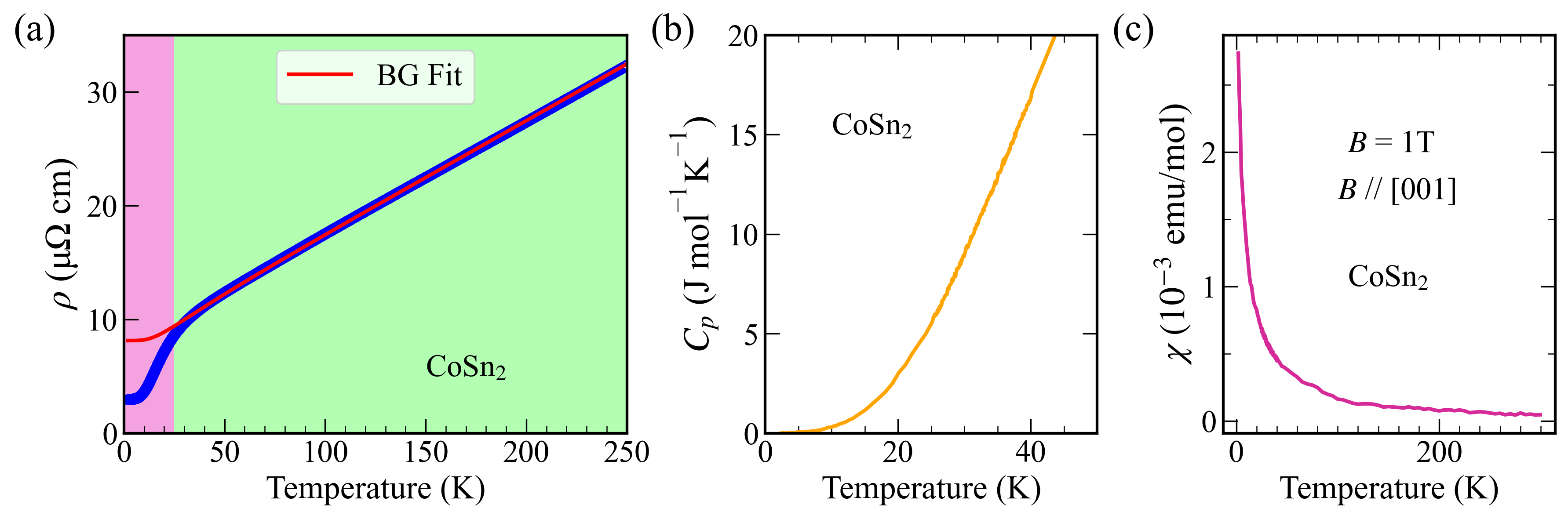}
		\caption{(a) Resistivity~($\rho$) as a function of temperature. The green region corresponds to the temperature range where the BG equation provides a good fit, while the purple region highlights the deviation from the fit. (b) Specific heat as a function of temperature. (c) $\chi$-$T$ curve of CoSn$_2$ measured in an applied field of 1~T, for $B~\parallel$~[001] direction.}
		\label{Figure5.3}
	\end{figure*}
	Figure~\ref{Figure5.3}(a) presents the temperature-dependent resistivity of ${\rm CoSn_2}$ measured in zero magnetic field, with the current applied along the [001] direction of the high temperature tetragonal phase~\cite{Armbruster}. While the resistivity exhibits typical metallic behavior, a noticeable slope change occurs below 25~K as indicated by purple color region in Fig.~\ref{Figure5.3}(a). When attempting to fit the data with Bloch-Gr\"{u}neisen (BG) expression~\cite{v.Minnigerode1983} 
	\begin{equation}
		\rho(T)=\rho_0+A_{ee}T^2+A_{ep} \left(\frac T\Theta_D\right)^5  \int_{0}^{\frac {\Theta_D} {T}} \frac{x^5 e^x}{(e^x -1)^2} dx
	\end{equation}
	\begin{figure*}[!ht]
		\centering
		\includegraphics[scale=0.5]{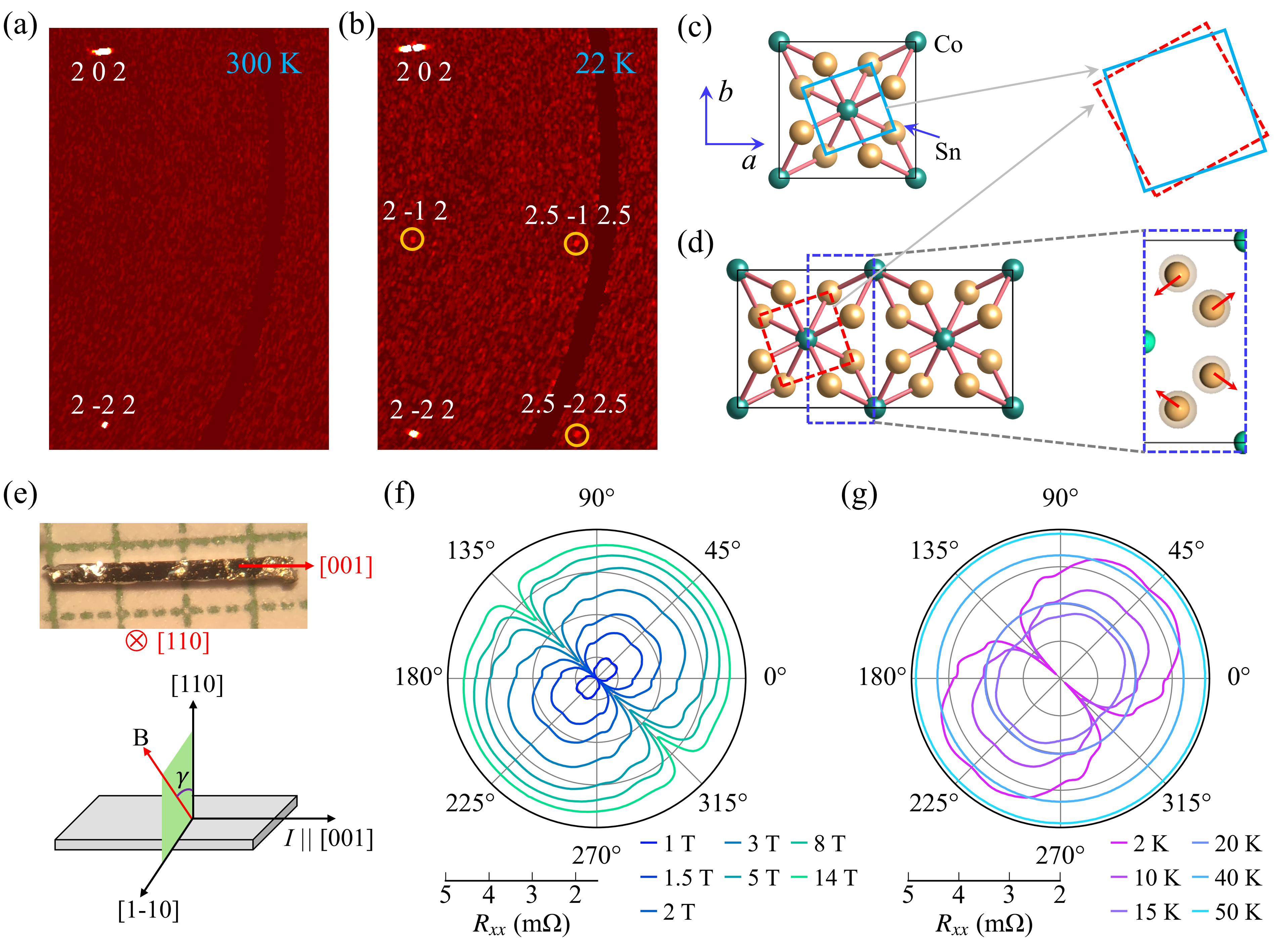}
		\caption{(a--b) Panels of reciprocal layer reconstructions of the $hkh$ plane at 300 and 22 K. Weak superlattice reflections are labeled by the broken Laue indices indicated by a yellow circle. At 22 K, there are also weak reflections like the 2 -1 2 that violates the $I$-centering. (c--d). View of the projections of the crystal structures at 300 and 22 K onto the \textbf{ab}  plane. The acentric $X$2 monoclinic lattice is a non-standard setting of the $C$2 structure where one observes a clear doubling of \textbf{a} and \textbf{c} respectively. Right panels illustrate the schematic distortion in a Sn atoms plane (between 300 and 22~K) and loss of 4-fold rotational symmetry at 22~K. Red arrows on the Sn atoms indicate the displacements away from the original positions. A detailed atomistic description is given in the SI \cite{SupplMat}. (e) As grown crystal and schematic of the sample geometry used for magnetic field rotation measurements. Polar plot of resistance for rotating magnetic field in $ab$ plane with $I~||~$[001] for (f) varying field at 2~K and (g) varying temperature at 3~T.}
		\label{Figure5.4}
	\end{figure*}
	
  where $\rho_0$ is the residual resistivity, $A_{ee}$ is electron-electron scattering coefficient, $A_{ep}$ is electron–phonon scattering coefficient and $\Theta_D$ is the Debye temperature, we observe a significant deviation from the fit at low temperatures. This deviation could be due to two possible factors: (i) a structural transition, or (ii) a magnetic transition, considering the presence of Co in the system. However, no noticeable anomaly are observed in the specific heat around the transition temperature with a high density of data points as shown in Fig.~\ref{Figure5.3}(b). We have also measured susceptibility ($\chi$) as a function of temperature at an applied field of 1~T. As shown in Fig.~\ref{Figure5.3}(c), the material exhibits paramagnetic behavior across the entire temperature range of 2–300~K, confirming that no magnetic transition is related to this slope change in resistivity. The absence of anomalies in these data left us puzzled about the origin of the effect.
  
  To investigate more carefully, the low temperature phase was examined by single-crystal x-ray diffraction (SXRD) through synchrotron radiation in the ID15B beamline at ESRF Grenoble France \cite{Garbarino2024-ID15B}. Analysis from Crysalis Pro software package and Jana 2006 \cite{crysalis, petricekv2014a} showed the crystal structure is described by the tetragonal symmetry $I$4/$mcm$ at ambient conditions. Cooling the crystal to 22 K revealed extremely weak satellite reflections at commensurate positions \textbf{q$^1$} = ($\frac{1}{2},0,\frac{1}{2}$) and \textbf{q$^2$} = ($0,\frac{1}{2},\frac{1}{2}$), which persists down to 10 K. \textbf{q$^2$} corresponds to satellite reflections from another domain related to the parent domain with \textbf{q$^1$} by the loss of the 4-fold rotational symmetry around \textbf{c*}. Figures~\ref{Figure5.4} (a--b) shows the reciprocal space reconstructions of the $hkh$ plane. Broken Laue indices correspond to weak superlattice reflections indicated by yellow circles for panel (b) that become more intense at 10 K as seen in the histograms in the SI \cite{SupplMat}. Additionally at 22 K, weak main Bragg reflections like 2 -1 2 violating $I$-center are also present coupled with a minute distortion of the lattice away from tetragonal symmetry as indicated by tiny degree of splitting of the reflection 2 0 2. This results in a doubling of the $a$ and $c$ axes, while $b$ remains as it is. Although the wavevector is commensurate, the superspace approach \cite{van2007incommensurate} was invoked. 
  
  This mathematical approach is used primarily to deal with incommensurate structures, where the intensities of the main Bragg reflections are kept distinct from the intensities of the satellites. However, it can be adapted to commensurate structures which would allow us to derive the superstructure from the parent basic structure by adopting a (3+1)-dimensional ($d$) setting. Using this approach both symmetries tetragonal $I$4/$mcm$ and its orthorhombic subgroup $Ibam$ are discarded as the wavevector \textbf{q} is incompatible \cite{stokesht2011a}. Initially the data was treated as incommensurate and subsequent refinements were done employing the superspace group $I$2/$c(\sigma_10\sigma_2$)0s. However, the model was discarded in favor of the commensurate approach where the supercell group is acentric monoclinic $X$2 $b$-unique which is a non-standard setting of $C$2. '$X$' represents user-defined centering translations to mimic the tetragonal setting as a $2a$$\times$$b$$\times2c$ lattice as shown in Figures.~\ref{Figure5.4}(c--d). Such non-standard settings are sometimes employed to allow for an easier comparison with the parent structure, for example in incommensurately modulated CuV$_2$S$_4$ the low temperature structure is described by the non-standard $F$-centering of acentric orthorhombic $Imm$2 to preserve the same setting as the parent cubic $Fd\bar{3}m$ lattice \cite{ramakrishnan2019a}. The supercell is chosen by employing $t_0 = \frac{3}{8}$ resulting in an improved fit of the superlattice reflections, thus confirming the absence of incommensurability. The specific value of $t_0 = \frac{3}{8}$ which represents discrete cuts or sections in the superspace is chosen as not all $t$ values can describe the modulation in the commensurate phase correctly. 
  
  While indexing as (3+1)-$d$, the basic structure parameters corresponding to the parent $I$4/$mcm$ structure appear pseudo-tetragonal, thereby suggesting a negligible distortion of the 3-$d$ lattice. Such negligible or absence of distortions of the lattice towards monoclinic symmetry is ubiquitous in many modulated systems namely Sm$_2$Ru$_3$Ge$_5$ \cite{bugaris2017charge}, SrAl$_4$ \cite{ramakrishnan2024a} and Rb$_2$ZnCl$_4$ \cite{kotla2025a}. For the final interpretation,  this was later transformed to 3-$d$ supercell $X$2 that is equivalent to monoclinic $C$2 $b$-unique in the standard setting, where the former is a 4-fold superstructure and the latter is a 2-fold as $C$2 possesses half the unit cell volume as $X$2. Although no distortion is visible to the naked eye when comparing left panels of Figs.~\ref{Figure5.4}(c) and \ref{Figure5.4}(d), the right panel of Fig.~\ref{Figure5.4}(c) presents a schematic representation of the distortion in the Sn square layer. The breaking of fourfold symmetry also becomes evident at 22~K (right panel of Fig.~\ref{Figure5.4}(d)), where the original structure is superimposed with its 90$^\circ$ rotated counterpart. The faint atoms represent the original Sn positions at 22 K, and red arrows indicate the
direction of displacement from the original to the rotated positions. All crystallographic information including lattice centering transformations, and description of the lowering of symmetry are given in the SI \cite{SupplMat}. 

Additional support for this symmetry lowering comes from angle dependent magnetoresistance measurements where the magnetic field is rotated within the $ab$ plane with the electrical current along the [001] direction (Fig.~\ref{Figure5.4}(e)). At temperatures above 40~K, the polar plots of resistance exhibit nearly isotropic behavior, consistent with the expected fourfold rotational symmetry of the high temperature tetragonal phase. However, upon cooling below 20~K, a clear two-fold symmetry emerges in the angular dependent resistance data, as shown in Fig.~\ref{Figure5.4}(g). When we increase the magnetic field at 2~K, as shown in Fig.~\ref{Figure5.4}(f), the two-fold symmetry persists for all field strengths. This transition from four-fold to two-fold symmetry corroborates the results of x-ray diffraction where the crystal symmetry reduces from $I$4/$mcm$ to $C$2. 

\begin{figure*}[!ht]
		\centering
		\includegraphics[scale=0.49]{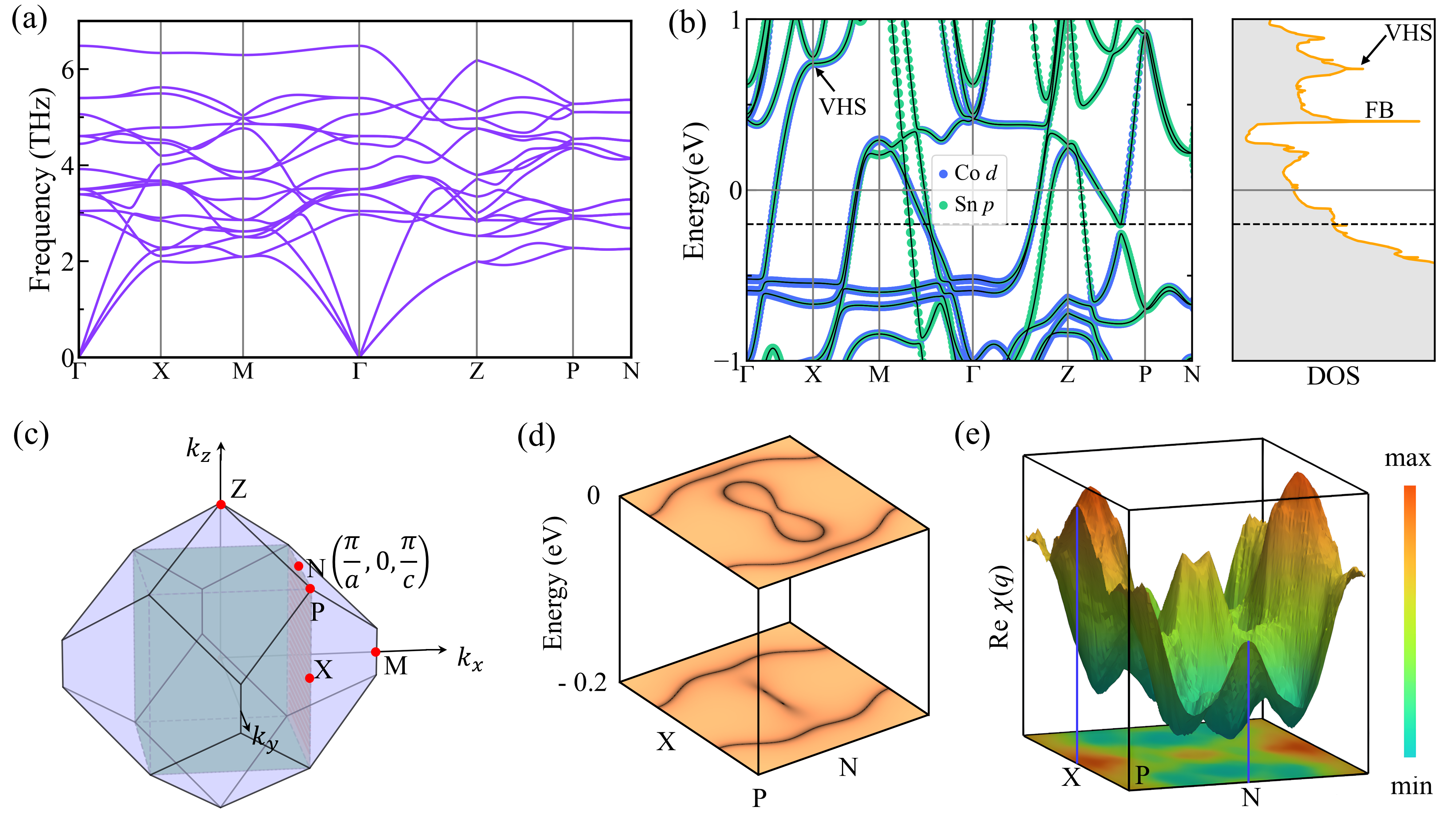}
		\caption{(a) Phonon band structure of CoSn$_2$. (b) Bulk band structure along with DOS. (c) Primitive and conventional Brillouin zones with high symmetry points marked; the red plane indicates the momentum slice used for further analysis. (d) Fermi surface contours at the indicated red plane plotted at two different energies. (e) Real part of the electronic susceptibility $\chi(q)$ in the same plane.}
		\label{Figure5.5}
	\end{figure*}
    \subsection{Origin of Symmetry breaking}
	To investigate the origin of lattice distortion, we calculated the phonon band dispersion of the high-temperature structure, as shown in Fig.~\ref{Figure5.5}(a). The phonon eigenvalues remain positive throughout the Brillouin zone (BZ), with no imaginary frequencies, indicating that the structure is dynamically stable. To examine the possibility of electronic instability, we have calculated electronic band structure along with Density of States (DOS) (Fig.~\ref{Figure5.5}(b)). For comparison with the ARPES measurements (discussed later), the Fermi level is shifted downward by $-0.2$ eV, as indicated by the dashed line. The presence of van Hove singularities (VHS) and nearly flat bands (FB) is observed, but they lie far from the Fermi level, suggesting that electronic correlations are not very strong in this system. Since superlattice reflections consistent with a $2\times1\times2$ modulation are detected in SXRD, we focus on the red plane in Fig.~\ref{Figure5.5}(c) of the primitive BZ, which is equivalent to $k_x=\pi/a$ plane of conventional BZ. On this plane, the electronic Fermi surface contours are computed both at the original Fermi energy and at the shifted value of $-0.2$ eV (Fig.~\ref{Figure5.5}(d)). Upon visual inspection, parallel contours separated by a finite wave vector are apparent, as shown in Fig.~\ref{Figure5.5}(d). This suggests potential nesting, prompting us to compute the bare charge susceptibility $\chi(q)$ using the Lindhard function~\cite{Dressel_Grüner_2002,PhysRevB.111.045135}
	\begin{equation}
		\chi(q)=-\sum_{k,m,n}^{} \frac{f(\epsilon_{n,k})-f(\epsilon_{m,k+q})}{\epsilon_{n,k}-\epsilon_{m,k+q}+i\gamma}
	\end{equation}
	where $\epsilon_{n,k}$ is the eigenvalue of band $n$-th at $k$, $f (\epsilon)$ is the Fermi-Dirac distribution, and $\gamma$ is an infinitesimal broadening parameter. For simplicity, we restricted the summation to the bands crossing the Fermi level, as they provide the dominant contribution to the susceptibility. The calculated real part of susceptibility $\chi(q)$ in the $k_x=\pi/a$ plane is shown in Fig.~\ref{Figure5.5}(e). If the maximum of the electronic susceptibility corresponding to the N and X points is large enough that can produce electronic instability. The presence of such instabilities at these points imply that a variety of superstructures could potentially emerge, depending on which nesting vectors are favored energetically. Importantly, these instabilities can cause a redistribution of the electron density along a preferred direction, breaking the fourfold rotational symmetry of the electronic structure down to twofold. Furthermore, this electronic instability can couple to the lattice which could induce microscopic lattice distortions. Such rotational symmetry breaking due to electronic instability, known as nematic order, has also been reported in other correlated electron systems~\cite{Yang2024,PhysRevB.91.155106}. To further explore this possibility, we performed ARPES measurements, which are discussed in the following section.
    \begin{figure*}[!ht]
		\centering
		\includegraphics[scale=0.48]{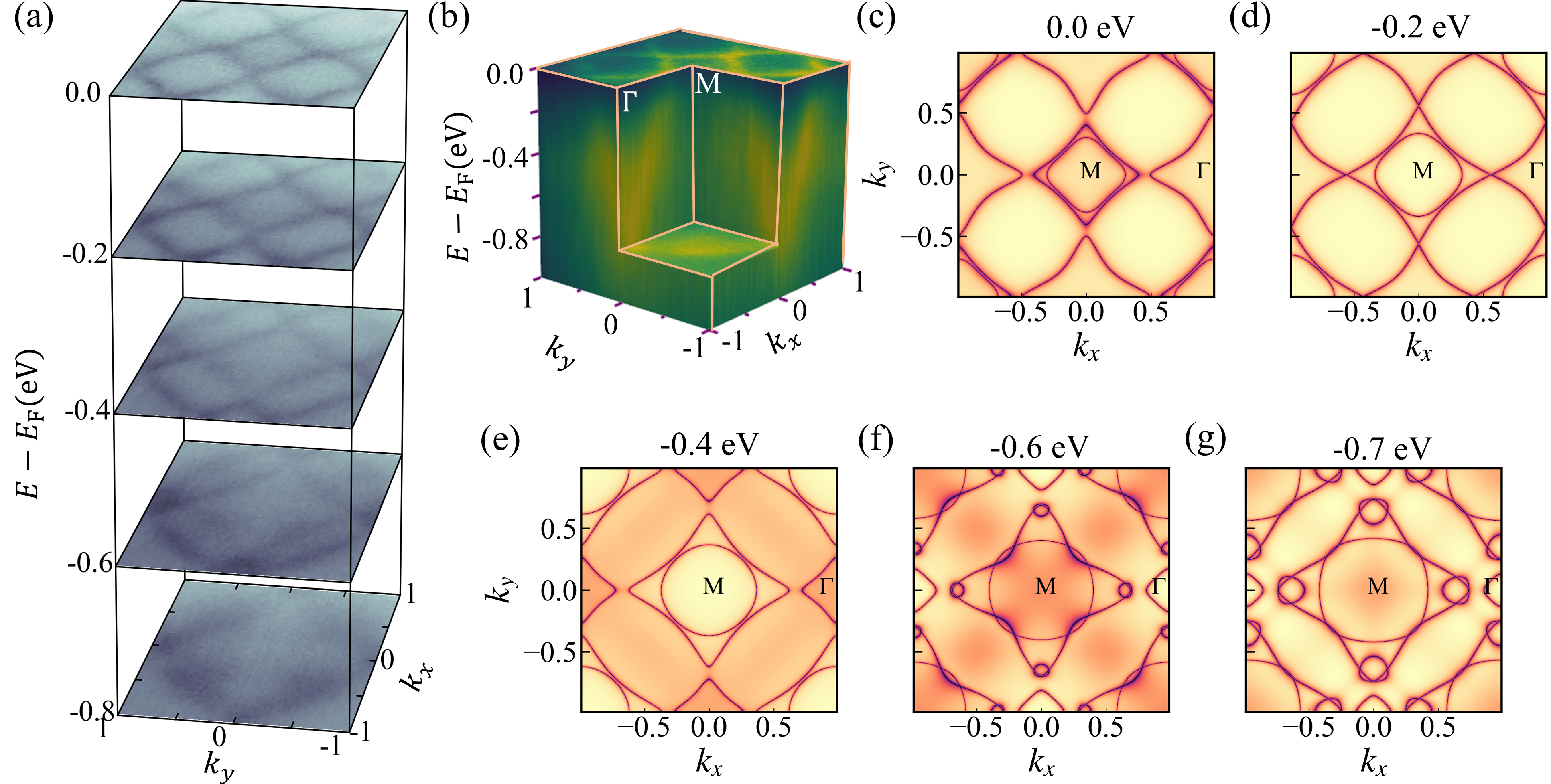}
		\caption{(a) Experimental Fermi surface maps at different binding energies. (b) 3D plot of the ARPES intensity map. (c-g) Calculated isoenergy contours at selected energies (0.0, -0.2, -0.4, -0.6, and -0.7 eV) in the $k_x$-$k_y$ plane.}
		\label{Figure5.7}
	\end{figure*}
	
	\subsection{ARPES}

	To investigate the three-dimensional electronic structure of CoSn$_2$, we performed ARPES measurements using a range of photon energies. Figure~\ref{Figure5.7}(a) displays constant energy contours at various binding energies, acquired with a photon energy of 128~eV at 20~K. The Fermi surface exhibits a diamond-shaped pocket surrounded by four lobes, forming a characteristic flower-like pattern. As the binding energy increases, the diamond-shaped pocket expands, and the lobes elongate along specific momentum directions, gradually evolving into more rectangular features. This evolution is further illustrated in the three dimensional ARPES intensity map shown in Fig.~\ref{Figure5.7}(b). This behavior closely matches our DFT calculations, assuming the M point is at the center of the Brillouin zone, as shown in Fig.~\ref{Figure5.7}(c-g). Notably, the inner pocket predicted by DFT is not observed in the ARPES measurements.

    \begin{figure*}[!ht]
		\centering
		\includegraphics[scale=0.52]{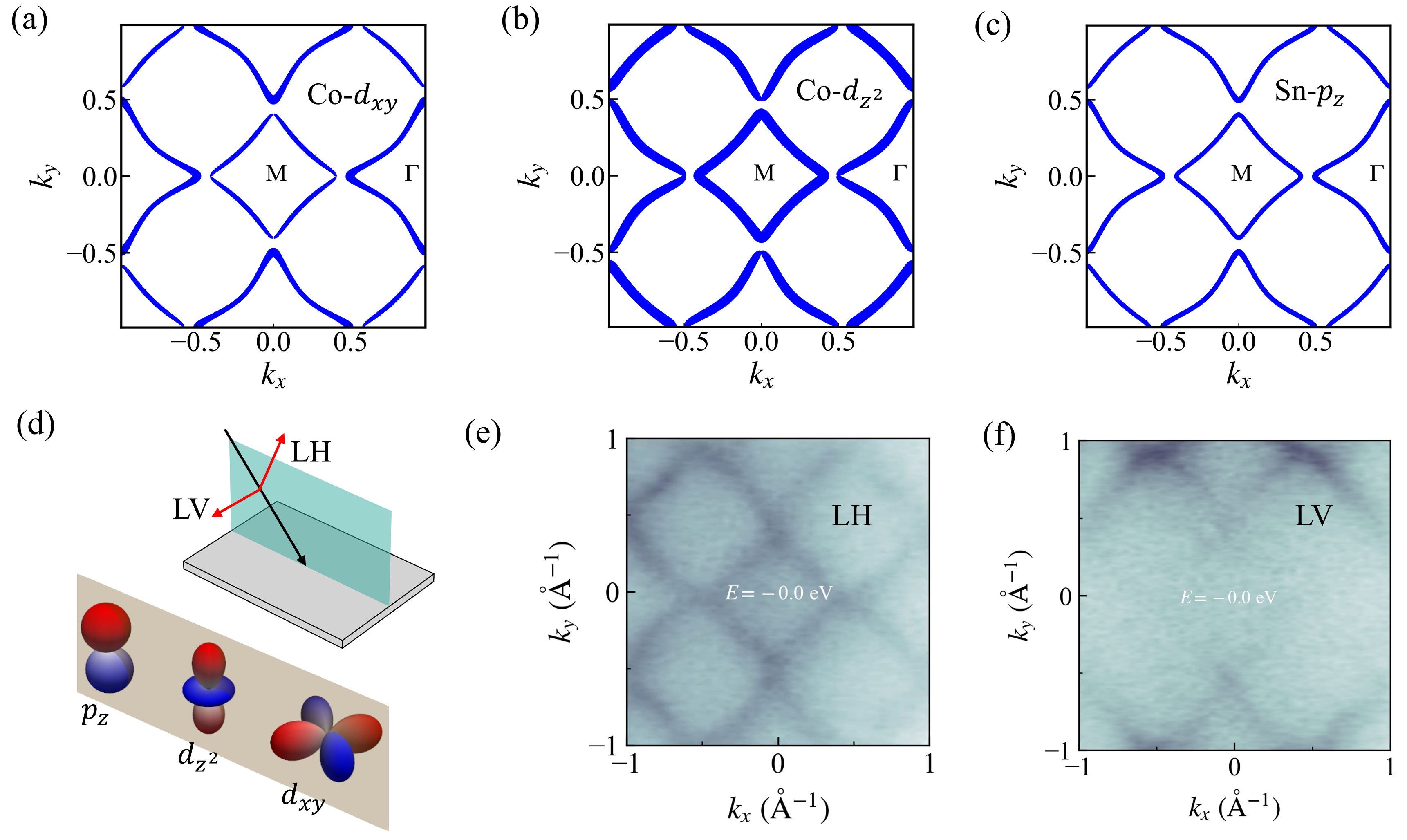}
		\caption{Calculated isoenergy contours at $E = E_F$ with specific orbital contributions: (a) Co-$d_{xy}$, (b) Co-$d_{z^2}$, and (c) Sn-$p_z$.  (d) Schematic of ARPES geometry showing linear horizontal (LH) and linear vertical (LV) polarizations relative to the sample surface, along with the spatial distribution of relevant orbitals ($p_z$, $d_{z^2}$, $d_{xy}$). (e) and (f) show ARPES constant energy maps at Fermi energy measured under LH and LV polarizations.}
		\label{Figure5.8}
	\end{figure*}
	Following the Fermi surface mapping, we explored the orbital symmetry characteristics using polarization dependent ARPES. Orbital-projected calculations along the observed Fermi contours reveal that Co-$d_{xy}$, Co-$d_{z^2}$, and Sn-$p_z$ orbitals contribute most prominently (Fig.~\ref{Figure5.8}(a-c)). To probe these contributions, we utilized both linear horizontal (LH) and linear vertical (LV) polarized photons. As illustrated in Fig.~\ref{Figure5.8}(d), the electric field vector of LH light lies within the mirror plane, while that of LV light is perpendicular to it. With respect to the mirror plane, the $d_{xy}$ orbital is even, whereas the $d_{z^2}$ and $p_z$ orbitals are odd. According to the photoemission selection rules, even-parity orbitals are primarily detected with LH polarization, while odd-parity orbitals are more visible with LV polarization. The Fermi surface map obtained with LH polarization in Fig.~\ref{Figure5.8}(e) exhibits a clear fourfold symmetry, reflecting the contributions from even-parity orbitals ($d_{z^2}$ and $p_z$) and preserving the crystal’s inherent fourfold symmetry. In contrast, the Fermi surface acquired under LV polarization (Fig.~\ref{Figure5.8}(f)) exhibits a significant intensity asymmetry, deviating strongly from the fourfold symmetry. This anisotropy may indicate a dominant contribution from the Co-$d_{xy}$ orbital, whose in-plane character can break the fourfold rotational symmetry of the electronic structure. While this reduction from fourfold to twofold symmetry in the ARPES intensity pattern is suggestive of orbital driven electronic nematicity, further investigation is necessary to firmly establish its origin.

\begin{figure*}[!ht]
		\centering
		\includegraphics[scale=0.47]{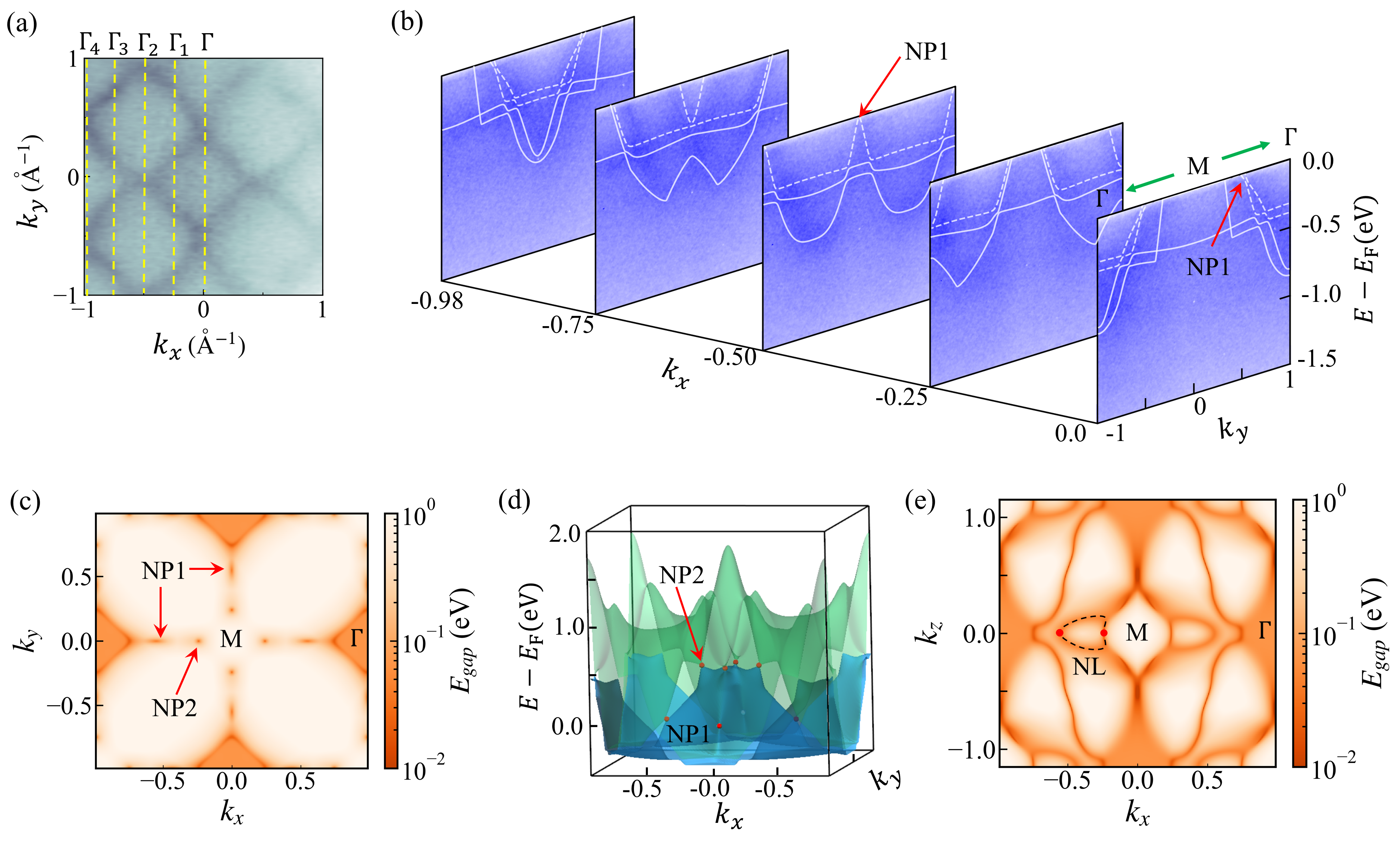}
		\caption{(a) ARPES intensity map at the Fermi level ($E_F$) in the $k_x$–$k_y$ plane, with vertical cuts ($\Gamma$–M direction) marked by dashed yellow lines. (b) ARPES intensity slices at successive $k_x$ planes from $k_x = -0.98$ to $k_x = 0$, showing the evolution of electronic bands along the $\Gamma$–M direction. Experimental bands matching with DFT calculations are overlaid in white. (c) Energy gap values between the white dashed bands in the $k_x$–$k_y$ plane, showing minimal band gap around band crossing or touching.
        (d) 3D band structure plot in the same plane, illustrating the anisotropic nature of NP1 with type-I and type-II crossings along perpendicular directions. 
        (e) Energy gap map in the $k_x$–$k_z$ plane, highlighting a continuous nodal loop (NL). NP1 and NP2 are part of this loop and are marked by red points on the dashed curve.}
		\label{Figure5.9}
	\end{figure*}

The band dispersion at multiple $k_x$ slices along the $\Gamma$--M direction is shown in Fig.~\ref{Figure5.9}(b). The observed bands that match with the DFT calculations are overlaid in white for visual clarity. A type-II band crossing is clearly visible at $k_x = 0$, while a type-I band crossing appears at $k_x = -0.5$~\AA~$^{-1}$ between two white dashed bands near the Fermi level (highlighted by red arrows). Although these crossings seem to occur at different $k_x$ positions, they actually correspond to the same point (NP1). This is confirmed by the energy gap values between these bands in $k_x$--$k_y$ presented in Fig.~\ref{Figure5.9}(c). The anisotropic nature of NP1 is further illustrated by its band dispersion in the $k_x$--$k_y$ plane (Fig.~\ref{Figure5.9}(d)), which reveals two distinct types of band crossings (type I and type II) along two perpendicular directions. In addition to NP1, another Dirac type crossing (NP2) is observed between the same two bands but located far from the Fermi level. While NP1 and NP2 may appear as isolated Dirac points in the band structure, they are in fact part of a continuous nodal loop (NL) lying in the $k_x$--$k_z$ plane. This nodal loop is illustrated in Fig.~\ref{Figure5.9}(e) as a dashed curve, with NP1 and NP2 corresponding to marked red points. Actually, in the presence of SOC, the nodal loop should not be strictly gapless. Although SOC is included in all the calculations and plots, there is only a very small gap opening, preserving the essential features of the nodal loop structure.

\section{Conclusions}\label{sec3}

In summary, our comprehensive investigation of CoSn$_2$ single crystals reveals a rich interplay between electronic, structural, and transport phenomena, despite the absence of magnetic ordering. A distinct slope change in resistivity at low temperatures is associated with the emergence of superlattice reflections that leads to a formation of a superstructure leading to a lowering of symmetry from tetragonal to acentric monoclinic. However, the absence of specific heat anomalies, the minimal nature of the distortion, and the lack of negative phonon frequencies all indicate that the crystal structure corresponding to the parent tetragonal lattice remains highly stable and resists deformation. Angle dependent magnetoresistance shows a clear four-fold to two-fold symmetry breaking below 25~K. But what drives this symmetry breaking in the absence of strong lattice distortions or magnetic ordering? To address this, we performed electronic susceptibility calculations, which suggest a strong tendency toward an electronically driven instability, capable of breaking the rotational symmetry of the electronic structure. Polarization dependent ARPES measurements further reveal orbital-selective anisotropy, may play the role in breaking rotational symmetry. Additionally, the observation of large linear magnetoresistance points toward the presence of Dirac-like carriers near the Fermi level, as also evidenced by ARPES measurements. Taken together, our findings position CoSn$_2$ as a compelling platform for studying symmetry breaking in an itinerant, nonmagnetic nodal line semimetal. This work broadens the landscape of nematic materials beyond correlated systems and underscores the need to explore similar phenomena in other high-symmetry transition metal compounds.

\section{Methods}\label{sec4}
\subsection*{Synthesis of CoSn$_2$ single crystals}
For growing single crystals of ${\rm CoSn_2}$ we have taken high purity  elements of Co (99.99\%;Alfa Aesar), and Sn (99.99\%;Alfa Aesar) in a clean alumina crucible with a specific molar ratio of 1:19. The crucible was sealed in a quartz ampoule under a partial argon atmosphere and heated in a resistive furnace. The sealed ampoule is gradually heated to 1050$^\circ$C at a rate of 50$^\circ$C/h and kept at this temperature for 48 hours for uniform melting. Then, the furnace is slowly cooled down to 300$^\circ$C at a rate of 2$^\circ$C/h, where we have used centrifugation method to remove excess Sn flux. Several tiny needle shape crystals are obtained. The elemental composition of the grown crystal  is determined by energy dispersive analysis by x-ray (EDX). 


\subsection*{Transport measurements}
The transport and specific heat measurements are conducted  in a 14~T physical properties measurement system~(PPMS, Quantum Design, USA), and magnetisation measurements are carried out using 7~T vibrating sample magnetometer~(SQUID VSM, Quantum Design, USA).
\subsection*{Single crystal x-ray diffraction and data processing}
Temperature dependent single-crystal x-ray diffraction were performed on the ID15B Beamline at the ESRF synchrotron in Grenoble, France. Data were obtained using $\omega$ scans
from -32$^\circ$ to 32$^\circ$ with 0.5$^\circ$ step width, employing radiation with a wavelength of $\lambda$  = 0.4099 \AA{} on the Eiger2X CdTe 9M PCD device detector. Intensities were integrated followed by scaling and empirical absorption correction with Laue symmetry 4/$mmm$
for the tetragonal phase and 2/$m$ for the superstructure phase in CrysAlis Pro software package \cite{crysalis}. The resulting reflection file was imported into Jana 2006 \cite{petricekv2014a} for structure refinements. For the superstructure phase two runs were processed. One with low flux to capture the strong main Bragg reflections and a second run with moderate-to-high flux by modifying the undulator gap to detect the weak superstructure reflections. Both runs were scaled together using the common reflections between them a scaling factor.

\subsection*{First-principles calculations}
The band structure and density of states are calculated using density functional theory (DFT) within Vienna \textit{ab initio} simulation~(VASP) package~\cite{PhysRevB.54.11169,PhysRevB.59.1758}. Generalized gradient approximation (GGA) in the form of Perdew-Berke-Erzndof (PBE) functional is utilized to incorporate exchange-correlation effects~\cite{PhysRevLett.77.3865}. A plane-wave kinetic energy cutoff of 375 eV is applied, and the Brillouin zone is sampled using an 11$\times$11$\times$11 Gamma grid of $k$-points. The Wannier90~\cite{MOSTOFI2008685} code is utilized to construct the tight-binding Hamiltonian based on Wannier orbitals. We use the WannierTools~\cite{WU2018405} package for post-processing, including the calculation of Fermi surface cross-sections at different energies and band dispersions along various directions. The phonon band structure is calculated using the finite displacement method as implemented in Phonopy~\cite{TOGO20151}, in conjunction with VASP. The electronic susceptibility is computed using an in-house Python code, employing a carefully converged fine $80 \times 80 \times 80$ $k$-point mesh within the bulk Brillouin zone and an $100 \times 100$ $q$-point mesh in the plane.

\subsection*{Angle-resolved photoemission spectroscopy measurements}
Single crystals of CoSn$_2$ were claved in ultrahigh vacuum with a base pressure better than 1.0 $\times$ 10$^{-10}$ mbar at 20~K. The spot size on the sample 
was 10 $\mu$m x 15  $\mu$m during ARPES measurements. A
Scienta Omicron DA30-L hemispherical electron energy analyzer was used for electron
detection. The sample was cooled to temperature below 20~K using closed cycle He cryostat. Various photon energies and light polarization was used. Total Energy resolution was varied between 30-33 meV. The angular resolution was $\approx 0.2^{ \circ}$ for all ARPES measurements.

\section*{Acknowledgements}

We acknowledge the Department of Atomic Energy (DAE) of the Government of India for financial support.
Beamtime at ID15B ESRF was awarded under the proposal no. BLC16319. We thank Jo$\tilde{\mathrm{a}}$o Elias F. S. Rodrigues for his technical assistance with data collection. Authors S.R. and P.R. thank the support from the Agence Nationale de la Recherche under the project SUPERNICKEL (Grant No. ANR-21-CE30-0041-04). We acknowledge the MAX IV Laboratory for the ARPES beamtime at BLOCH. Research conducted at MAX IV, a Swedish national user facility, is supported by the Swedish Research Council under Contract No. 2018-07152, the Swedish Governmental Agency for Innovation Systems under Contract No. 2018-04969, and Formas under Contract No. 2019-02496.

\bibliography{ref}

\clearpage
\newpage

\begin{center}
	\textbf{Supplemental Information\\[24pt]
Evidence of electronic instability driven structural distortion in the nodal line semimetal CoSn$_2$}\\[12pt]
	Suman Nandi	$^{1}$, Bishal Maity $^{1}$, Shovan Dan	$^{1}$, Khadiza Ali $^{2}$, Bikash Patra	$^{1, 3}$, Anshuman Mondal	$^{4}$,  Gaston Garbarino	$^{4}$, Pierre Rodi\`ere $^{5}$, Sitaram Ramakrishnan $^{5}$, Bahadur Singh~$^{1}$, and Arumugam Thamizhavel $^{1}$\\[.2cm]
	{\itshape
		$^{1}$Department of Condensed Matter Physics and Materials Science, Tata Institute of Fundamental 	Research, Homi Bhabha Road, Colaba, Mumbai 400005, India\\
	$^{2}$MAX IV Laboratory, Lund University, Lund, SE-221 00, Sweden\\
    $^{3}$Department of Physical and Applied Sciences, Indian Institute of Information Technology, Surat, Gujarat
394190, India\\
	$^{4}$ESRF -- European Synchrotron Radiation Facility, 38000 Cedex Grenoble, France\\
    $^{5}$Institut N\'{e}el CNRS/UGA UPR2940, 25 Rue des Martyrs, 38042 Grenoble, France
	}
	(Dated: \today)
	\\[1cm]
\end{center}
Supplemental Information for CoSn$_2$:\\[12pt]
S1.\hspace{6mm} Symmetry of the superstructure phase at 22 K.\\
  S2.\hspace{6mm} Transformation from the X-centered to the C-centered setting.\\
 S3.\hspace{6mm} Crystallographic tables.\\
 S4.\hspace{6mm} Atomistic description of the loss of 4-fold and inversion symmetry.\\
S5.\hspace{6mm} Magnetoresistance.\\
S6.\hspace{6mm} Photon energy dependent ARPES.\\

\setcounter{equation}{0}
\renewcommand{\theequation}{S\arabic{equation}}
\setcounter{figure}{0}
\renewcommand{\thefigure}{S\arabic{figure}}
\setcounter{section}{0}
\renewcommand{\thesection}{S\arabic{section}}
\setcounter{table}{0}
\renewcommand{\thetable}{S\arabic{table}}
\setcounter{page}{1}
\renewcommand{\thesubsection}{{S\arabic{section}.\arabic{subsection}}}
\renewcommand{\thesubsubsection}{{S\arabic{section}.\arabic{subsection}.\arabic{subsubsection}}}
\renewcommand\baselinestretch{1.1}
\renewcommand{\topfraction}{0.99}
\renewcommand{\bottomfraction}{0.5}
\renewcommand{\textfraction}{0.0}
\renewcommand{\floatpagefraction}{0.8}
\setlength{\tabcolsep}{5pt}

\clearpage
\section{Symmetry of the superstructure phase at 22 K}

Processing of the data sets
collected at 300, 22 and 10 K where
done with the software CrysAlis Pro
\cite{crysalis}. For all temperatures except at 10 K (partial run)
this resulted in lattice parameters,
and a list of integrated intensities of Bragg reflections.
At this point, no deviations from tetragonal
symmetry could be detected from the basic structure.
Structure refinements have been performed
with the software {Jana2006} against the
integrated intensities of Bragg
reflections \cite{petricekv2014a}.

At 22 K with the unit cell parameters $a$ = 6.3407(17), $b$ = 6.349(6), $c$ = 5.4405(9), $\alpha$ = $\beta$ = 
$\gamma$ = 90$^\circ$ \;, satellite reflections appear at commensurate positions \textbf{q} = $(\frac{1}{2}, \frac{1}{2}, \frac{1}{2})$ as shown in Figure. However, on the basis of indexing, the modulation
wavevectors are found out to be \textbf{q$^1$} = $(\frac{1}{2}, 0, \frac{1}{2})$ and \textbf{q$^2$} = $(0, \frac{1}{2}, \frac{1}{2})$ where both \textbf{q$^1$} and \textbf{q$^2$} belong to two different twin domains and are related by the loss of four-fold rotational symmetry around \textbf{c*} as shown in equation \ref{rotation}.

\begin{figure}[ht]
\centering
\includegraphics[width=160mm]{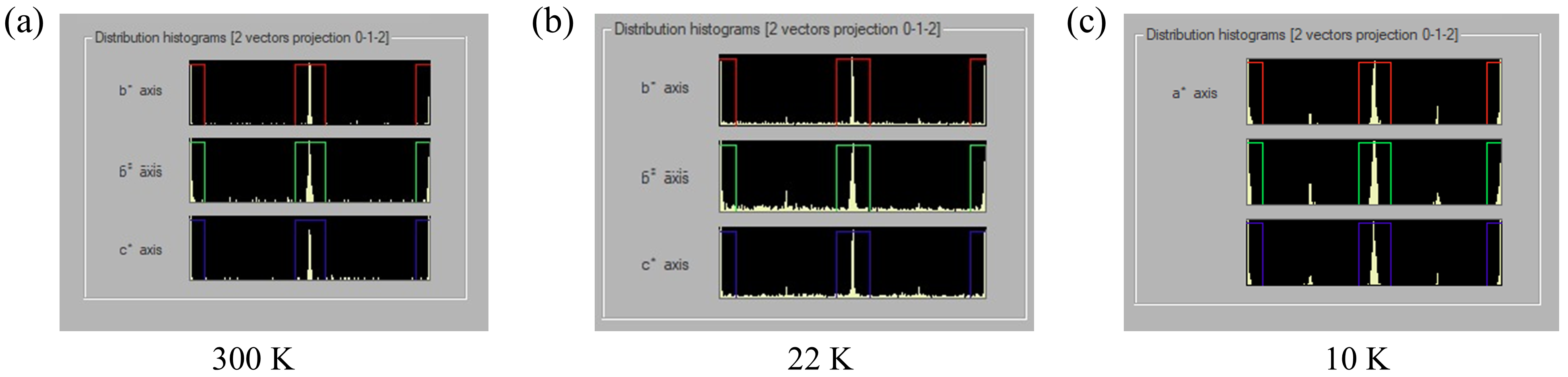}
\caption{(a--c) Histograms indicating the appearance of superlattice reflections at $\frac{1}{2}, \frac{1}{2}, \frac{1}{2}$ at lower temperatures. In reality they are \textbf{q$^1$} = $(\frac{1}{2}, 0, \frac{1}{2})$ and \textbf{q$^2$} = $(0, \frac{1}{2}, \frac{1}{2})$ belonging to two domains related by the loss of 4-fold rotation. Superlattice reflections appear more intense at 10 K.}
\label{FigureS1}
\end{figure}

\begin{equation}
\label{rotation}
\textbf{q$^2$} =
\begin{pmatrix}
                         0 & -1 & 0 \\
                         1 & 0 & 0 \\
                         0 & 0 & 1
                       \end{pmatrix}
                       \textbf{q$^1$} = 
                       \begin{pmatrix}
                         0 & -1 & 0 \\
                         1 & 0 & 0 \\
                         0 & 0 & 1
                       \end{pmatrix}\left(
                       \begin{array}{c}
 \frac{1}{2}  \\
0 \\
 \frac{1}{2}
\end{array} \right)=
\left( \begin{array}{c}
 0  \\
 \frac{1}{2}          \\
  \frac{1}{2} \\
\end{array} \right)
\end{equation}
\\

This results in a doubling of the basis vectors \textbf{a} and \textbf{c} in direct space respectively.
To resolve the symmetry of the superstructure we employed the superspace approach \cite{van2007incommensurate} by keeping the lattice in a (3+1)-d setting by describing the position of the superlattice reflections with respect to the main Bragg reflections by using the modulation wavevector \textbf{q} and integrating the intensities, keeping them distinct. This method is advantageous as it allows us to preserve the same unit cell volume as the high temperature tetragonal ($I$4/$mcm$) structure which is less cumbersome to deal with as opposed to integrating it as a supercell. 

The displacement modulation is described by
a modulation function for each atom of the form
\begin{equation}
\mathbf{u}(\bar{x}_{s4}) =
(u_{x}(\bar{x}_{s4}),\,u_{y}(\bar{x}_{s4}),\,
u_{z}(\bar{x}_{s4})) \, ,
\label{eqsm:sral4_modulation_function_def}
\end{equation}
where for $\alpha = x, y, z$,
\begin{equation}
u_{\alpha}(\bar{x}_{s4}) = \sum_{n=1}^{n_{max}}\,\{
A_{n,\alpha} \sin(2\pi n\bar{x}_{s4}) +
B_{n,\alpha} \cos(2\pi n\bar{x}_{s4}) \} \, .
\label{eqsm:sral4_modulation_function_harmonics}
\end{equation}
We have used $n_{max} = 1$ (up to first-order harmonics).

Both tetragonal $I$4/$mcm$ and its orthorhombic subgroup $Ibam$ cannot
be considered as they are incapable of incorporating the wavevector
\textbf{q} = $(\frac{1}{2}, 0, \frac{1}{2})$ \cite{stokesht2011a}. We therefore reduce it to monoclinic symmetry and a reasonable fit to the data employing the commensurate
approach by choosing the supercell group $X$2 of the superspace group $I$2/$c(\sigma_1\,0\,\sigma_2)0s$ with $t_0$ = $\frac{3}{8}$. $t_0$ represents discrete cuts or sections of the incommensurate structures superspace. For a more detailed description refer to \cite{van2007incommensurate}. The 'X'-centered setting is non-standard defined by user defined centering
translations to closely resemble the high temperature structure as it is no longer
possible to describe the lattice as $I$-centered.  For the $X$2, $3d$ supercell, there are 3 crystallographic positions for the Co and 4 crystallographic positions for Sn. From the atomic coordinates of Sn it is evident that the center of inversion is lost. We have also transformed from
$X$2 to $C$2, where the latter is now in the standard setting as discussed in Section \ref{sec:cosn2_transform}. Crystallographic details are listed in Tables \ref{tab:cosn2_superstructure} and \ref{tab:cosn2_compare}. The CIF of the $I$4/$mcm$ phase at 300 K as well as the $X$2 and equivalent $C$2 superstructures at 22 K are provided. The commensurate modulation amplitudes are not discussed as there are only 40 unique observable superlattice reflections despite the use of high flux, signifying the weak nature of the transition towards the superstructure. Nevertheless, we have provided the CIF of the structure in (3+1)$d$ which is equivalent to the 3$d$ supercell $C$2. Note that for non-standard settings it will be impossible to plot the crystal structures properly unless you describe the centering vectors for the $X$2 centering as listed in Table \ref{tab:cosn2_compare}.

\section{\label{sec:cosn2_transform}
Transformation from the $X$-centered to the $C$-centered lattice}

At 22 K  CoSn$_2$ undergoes a doubling of both the \textbf{a} and
\textbf{c}. This results in a 4-fold superstructure and the symmetry
is acentric monoclinic $X$2. '$X$' is a setting that is defined
by user defined centering vectors as described in Table \ref{tab:cosn2_compare}.
in order to make it easier to relate to the parent tetragonal structure.
However, in order to derive the standard setting $C$2 we transform the
the basis vectors $\mathbf{a}_{X}, \mathbf{b}_{X}, \mathbf{c}_{X}$ in the X-centered
setting to the basis vectors $\mathbf{a}_{X}, \mathbf{b}_{X}, \mathbf{c}_{X}$ in the $C$-centered setting by employing the transformation matrix derived from Jana 2006 \cite{petricekv2014a}.

\begin{equation}
\left( \begin{array}{c}
\mathbf{a}_{C} \\
\mathbf{b}_{C}  \\
\mathbf{b}_{C}
\end{array} \right)=
Q\textsuperscript{-1,t} \left( \begin{array}{c}
\textbf a_X  \\
\textbf b_X  \\
\textbf c_X
\end{array} \right)=
\begin{pmatrix}
                         -3/2 & 0 & -1/2 \\
                         0 & -1 & 0 \\
                         1/2 & 0 & 1/2
                       \end{pmatrix} \left(
                       \begin{array}{c}
\textbf a_X  \\
\textbf b_X  \\
\textbf c_X
\end{array} \right)=
\left( \begin{array}{c}
 -\frac{3}{2}(\mathbf{a}_{X})  -\frac{1}{2}(\mathbf{c}_{X})  \\
 -\mathbf{b}_{X}          \\
 \frac{1}{2}(\mathbf{a}_{X}) + \frac{1}{2}(\mathbf{c}_{X}) \\
\end{array} \right)
\end{equation}
\\

\begin{equation}
\begin{array}{c l}
\mathbf{a}_{C} &= -\frac{3}{2}(\mathbf{a}_{X})  -\frac{1}{2}(\mathbf{c}_{X}) \\
\mathbf{b}_{C} &= -\mathbf{b}_{X} \\
\mathbf{c}_{C} &= \frac{1}{2}(\mathbf{a}_{X}) + \frac{1}{2}(\mathbf{c}_{X}) \\
\end{array}
\label{eqsm:cosn2_x_to_c_center}
\end{equation}

For the $X$-centered orthorhombic lattice we have the lattice
parameters $a$ = 12.6814, $b$ = 6.3490, $c$ =  10.8810, $\alpha$ = $\beta$ = $\gamma$ = 90.
Thus by taking dot products of the basis vectors with itself we have:

\begin{equation*}
\begin{array}{c l}
{a}_{C} &= \sqrt{(-\frac{3}{2}{a}_{X})^2 + (-\frac{1}{2}{c}_{X})^2} =
\sqrt{(-\frac{3}{2}*12.6814)^2 + (-\frac{1}{2}*10.8810)^2} = 19.7848 \; \mathrm{\AA} \\
{b}_{C} &= {b}_{X} = 6.3940 \; \mathrm{\AA} \\
{c}_{C} &= \sqrt{(\frac{1}{2}{a}_{X})^2 + (\frac{1}{2}{c}_{X})^2}
= \sqrt{(\frac{1}{2}*12.6814)^2 + (\frac{1}{2}*10.8810)^2} = 8.3548 \; \mathrm{\AA} \\
\end{array}
\label{eqsm:cosn2_eq1}
\end{equation*}

To compute $\beta_C$ we now take the inverse of the matrix Q such that we have:

\begin{equation}
\left( \begin{array}{c}
\mathbf{a}_{X} \\
\mathbf{b}_{X}  \\
\mathbf{b}_{X}
\end{array} \right)=
Q\textsuperscript{t} \left( \begin{array}{c}
\textbf a_C  \\
\textbf b_C  \\
\textbf c_C
\end{array} \right)=
\begin{pmatrix}
                         -1 & 0 & -1 \\
                         0 & -1 & 0 \\
                         1 & 0 & 3
                       \end{pmatrix} \left(
                       \begin{array}{c}
\textbf a_C  \\
\textbf b_C  \\
\textbf c_C
\end{array} \right)=
\left( \begin{array}{c}
 -(\mathbf{a}_{X})  + -(\mathbf{c}_{X})  \\
 -\mathbf{b}_{X}          \\
 (\mathbf{a}_{X}) + 3(\mathbf{c}_{X}) \\
\end{array} \right)
\end{equation}
\\

\begin{equation}
\begin{array}{c l}
\mathbf{a}_{X} &= -(\mathbf{a}_{C})  + -(\mathbf{c}_{C}) \\
\mathbf{b}_{X} &= -\mathbf{b}_{C} \\
\mathbf{c}_{X} &=  (\mathbf{a}_{C}) + 3(\mathbf{c}_{C})  \\
\end{array}
\label{eqsm:cosn2_x_to_c_center}
\end{equation}

As $\beta_C$ is the angle between $\mathbf{a}_{C}$ and $\mathbf{c}_{C}$ we now take the dot product of \ref{eqsm:dotprod2}
with itself.

\begin{equation}
\begin{array}{c l}
\mathbf{c}_{X} &=  (\mathbf{a}_{C}) + 3(\mathbf{c}_{C}) \\
\end{array}
\label{eqsm:dotprod2}
\end{equation}

Now we have the square of the the lattice parameter $c_{X}$:

\begin{equation*}
\begin{array}{c l}
c_{X}^2 = (a^2_C + (3c)^2_C + 6a_Cc_C\cos\beta_C) \\
\end{array}
\end{equation*}

\begin{equation*}
\begin{array}{c l}
\beta_C = \cos(\frac{c_{X}^2 - a^2_C - (3c)^2_C}{6a_Cc_C})^{-1} = 155.13^{\circ} \\
\end{array}
\end{equation*}

Finally $\beta_C$ is computed but keep in mind that $\beta_X$ in the $X$-centered setting is
 90$^\circ$ and that the monoclinic distortion is negligible.
The large deviation of $\beta_C$ from 90$^\circ$ is just a result of transformation
to the standard $C$-centered setting.
\begin{figure*}[!ht]
		\centering
		\includegraphics[scale=0.5]{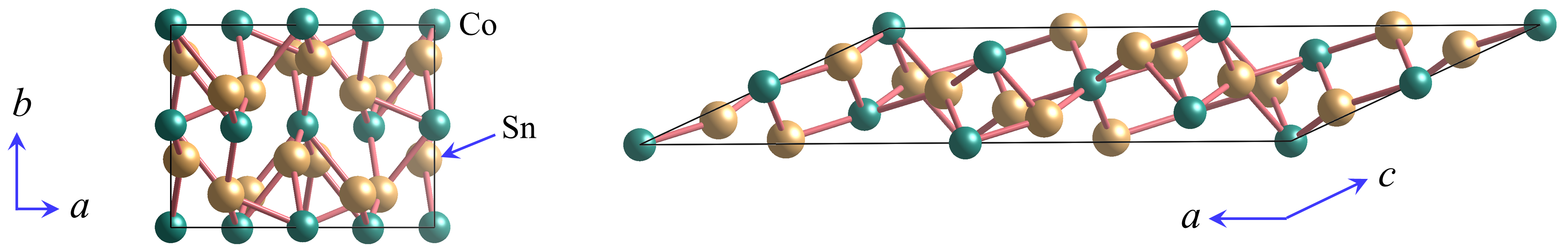}
		\caption{Projections of the crystal structure at 22~K onto the \textbf{ab} and \textbf{ac} planes in the $C$2 setting. The angle between \textbf{a} and \textbf{c} is 155.13$^{\circ}$.}
		\label{FigureS6}
	\end{figure*}

\clearpage

\section{Crystallographic tables}

\begin{table}[!ht]
\caption{\label{tab:cosn2_superstructure}%
Crystallographic data of crystal A of
CoSn$_2$ at 300 K and 22 K (superstructure phase). Note that some of statistics especially the fit
to the satellites is not optimal due its very weak intensity despite the use of high flux.}
\small
\begin{ruledtabular}
\begin{tabular}{ccc}
Temperature (K) & 300      & 22  \\
Modulation phases & -   &  $t_{0} = \frac{3}{8}$     \\
Space/Superspace group & $I4/mcm$  &
$I$2/$c(\sigma_1\,0\,\sigma_2)0s$ \\
 No. (cite) & 140 & {15.1.4.1}   \\
$a$ (\AA{}) & 6.3635(10)  & 6.3407(17) \\
$b$ (\AA{}) & 6.3635  & 6.349(6) \\
$c$ (\AA{}) & 5.4523(18) & 5.4405(9)  \\
Volume (\AA{}$^3$) & 220.79(9) & 219.0(2)  \\
Wavevector \textbf{q} & --
& $\frac{1}{2}$\textbf{a*}+$\frac{1}{2}$\textbf{c*} \\
$Z$ & 4 & 4 \\
Wavelength (\AA{}) & 0.4099 & 0.4099  \\
Detector distance (mm) &170.85 &  170.85  \\
$\omega$-scans (deg) &-32 to +32 & -32 to +32 \\
Rotation per image (deg) & 0.5 & 0.5 \\
$(\sin(\theta)/\lambda)_{max}$ (\AA{}$^{-1}$) &0.879387& 0.873974\\
Absorption, $\mu$ (mm$^{-1}$) & 30.230 & 30.474  \\
T$_{min}$, T$_{max}$ & 0.81, 1.82 & 0.70. 1.60  \\
Criterion of observability & $I>3\sigma(I)$ & $I>3\sigma(I)$  \\
No. of reflections measured, \\
$(m = 0)$  &  312  & 244 \\
$(m = 1)$  & -     & 592 \\
No. of unique reflections,   \\
$(m = 0)$ (obs/all) & 99/119 & 124/172  \\
$(m = 1)$ (obs/all) & -- & 40/231  \\
$R_{int}$ $(m = 0)$ (obs/all) &0.0181/0.0182 & 0.0181/0.0182 \\
$R_{int}$ $(m = 1)$ (obs/all) &-- & 0.0881/0.3544 \\
No. of parameters &8 & 42 \\
$R_{F }$ $(m = 0)$  (obs) &0.0633 & 0.0763\\
$R_{F }$ $(m = 1)$ (obs) &- & 0.1742 \\
$wR_{F }$ (all)  &0.1573 & 0.1075\\
GoF (obs/all) &7.17/10.94 & 6.53/3.95\\
$\Delta\rho_{min}$, $\Delta\rho_{max}$(e \AA$^{-3}$) &
 -3.62, 3.93 & -3.68, 3.04  \\
\end{tabular}
\end{ruledtabular}
\end{table}

\begin{table}[ht]
\caption{\label{tab:cosn2_compare}%
Crystallographic data of crystal A of CoSn$_2$ upon transformation to the supercell
 at 22 K. Two settings are shown which are both equivalent. One is $X$2 where '$X$' corresponds
 to user defined centering translations which are designed to mimic the setting of the high
 temperature phase. We also show the transformed cell in the standard $C$-centering adopting
 the acentric monoclinic space group $C$2. Note the statistics are the same as they are equivalent. Only the scale has been refined after transforming to $3d$ from $(3+1)d$.}
\small
\begin{ruledtabular}
\begin{tabular}{ccc}
Space group & $X$2  &
$C$2\\
Centering vectors &  (0.75  0.5  0.25), (0.5  0  0.5), (0.25  0.5  0.75)   & (0.5, 0.5 0) \\           
 No. & 15 & 15   \\
$a$ (\AA{}) & 12.6814(17)  & 19.7833(17) \\
$b$ (\AA{}) & 6.349(6)  & 6.349(6) \\
$c$ (\AA{}) &  10.8810(9) & 8.3536(9)  \\
$\beta$ ($^{\circ}$) & 90.02(2) & 155.32(3)  \\
Volume (\AA{}$^3$) & 876.1(2) & 438.1(2)  \\
$Z$ & 16 & 8 \\
Criterion of observability & $I>3\sigma(I)$ & $I>3\sigma(I)$  \\
No. of unique reflections,   \\
(obs/all) & 201/496 & 201/496  \\
$R_{F }$   (obs) &0.0739 & 0.0739\\
$wR_{F }$ (all)  &0.1030 & 0.1030\\
GoF (obs/all) &5.31/3.50 & 5.31/3.50\\
$\Delta\rho_{min}$, $\Delta\rho_{max}$(e \AA$^{-3}$) &
 -3.6, 3.03 & -3.68, 3.07  \\
\end{tabular}
\end{ruledtabular}
\end{table}

\section{Atomistic description of the loss of 4-fold and inversion symmetry}
\begin{figure*}[!ht]
		\centering
		\includegraphics[scale=0.5]{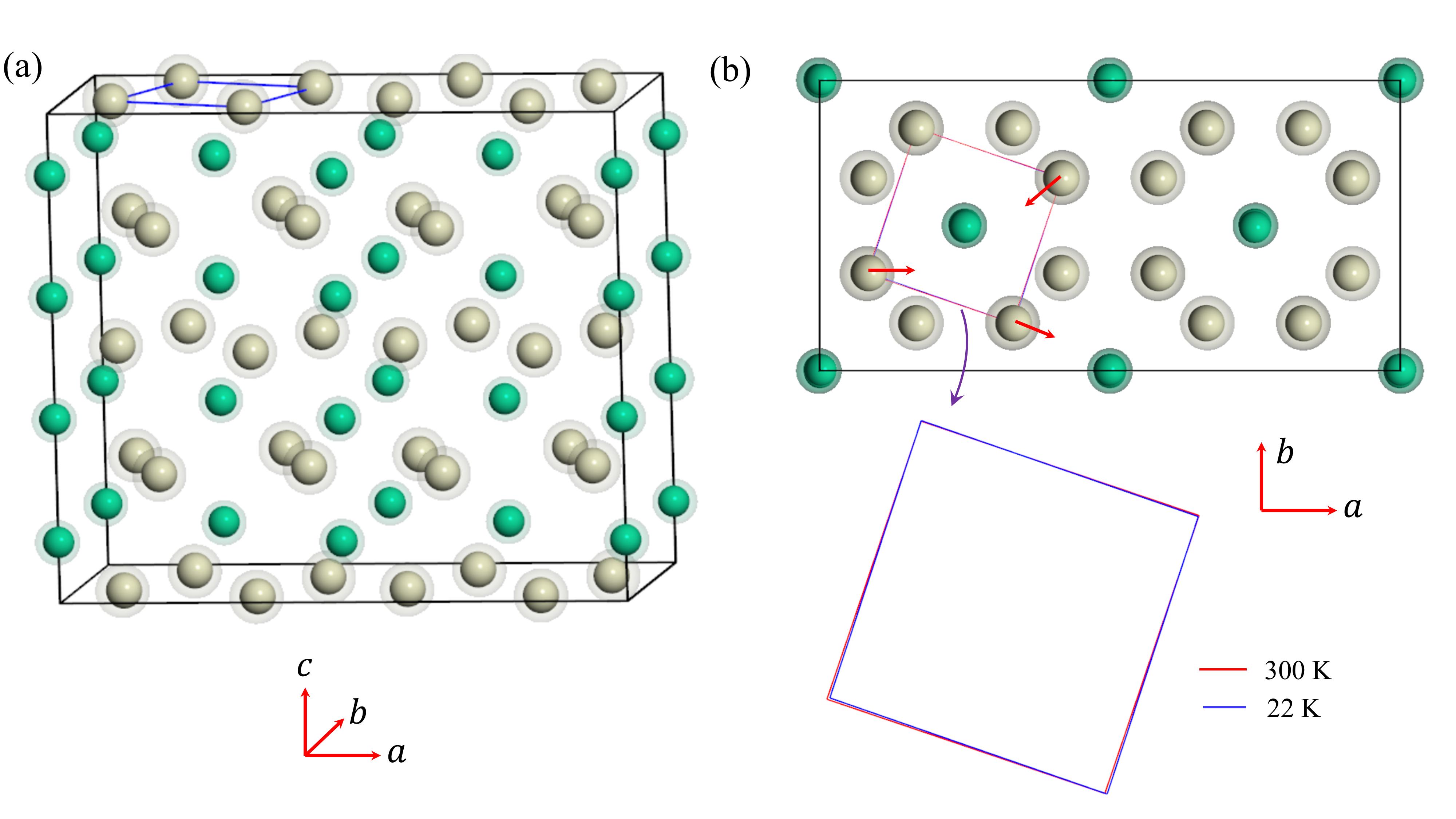}
		\caption{Comparison of the 300~K and 22~K structures. (a) Side view showing the structural arrangement. (b) Top view highlighting the local displacement (red arrows) of Sn atoms. The lower inset shows the distortion of a square formed by four Sn atoms in one layer, with outlines at 300~K (red) and 22~K (blue).}
		\label{FigureS2}
	\end{figure*}

Figure \ref{FigureS2}(a) compares the crystal structure of CoSn$_2$ at 300~K (shown as faint/shadowy atoms) and 22~K (shown as solid atoms). The 300~K structure has been expanded by a factor of $2 \times 1 \times 2$ to match the unit cell dimensions of the 22~K structure for direct comparison. A careful inspection of the projection along the $c$-axis reveals a very slight distortion in atomic positions between the two temperatures, as shown in Fig.~\ref{FigureS2}(b). This distortion is indicated by red arrows that shows the displacement of Sn atoms from their 300~K to 22~K positions. To highlight this subtle change, we compare a square formed by the top layer of four Sn atoms at both temperatures, which depicts a minute lattice distortion occurring at low temperatures.
\begin{figure*}[!ht]
		\centering
		\includegraphics[scale=0.5]{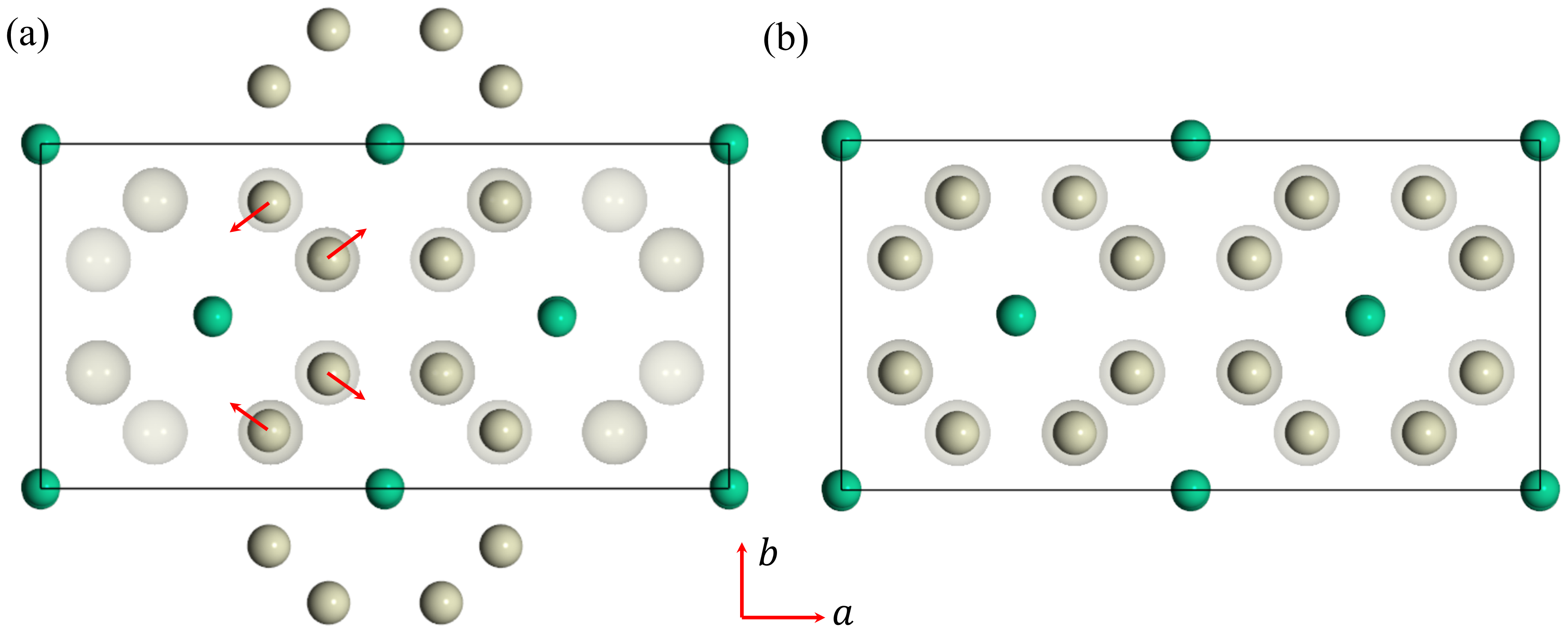}
		\caption{(a) In the 22~K structure, Sn atoms are rotated by 90$^\circ$, breaking the four-fold rotational symmetry. The faint atoms represent the original Sn positions at 22~K. Red arrows indicate the direction of displacement from the original to the rotated positions. (b) A 180$^\circ$ rotation of the Sn atoms preserves the two-fold symmetry of the structure.}
		\label{FigureS3}
	\end{figure*}

Figure ~\ref{FigureS3} illustrates the effect of atomic displacements of CoSn$_2$ at low temperature. In Fig ~\ref{FigureS3}(a), the faint atoms represent the original structure at 22~K, while the solid atoms show the same structure rotated by 90$^\circ$. Four-fold rotational symmetry would lead to perfect overlap of the rotated structure with the original. However, the misalignment between the faint and solid Sn atoms, highlighted by the red arrows, indicates that the four-fold symmetry is broken. This is in contrast to Fig ~\ref{FigureS3}(b) that shows a 180$^\circ$ rotation of the same structure. Here, the rotated and original atomic positions align exactly, confirming that two-fold rotational symmetry is preserved. It is realized that the low-temperature structure of CoSn$_2$ breaks four-fold symmetry due to minor displacements in the Sn sub-lattice, while possessing two-fold symmetry.

\begin{figure*}[!ht]
		\centering
		\includegraphics[scale=0.5]{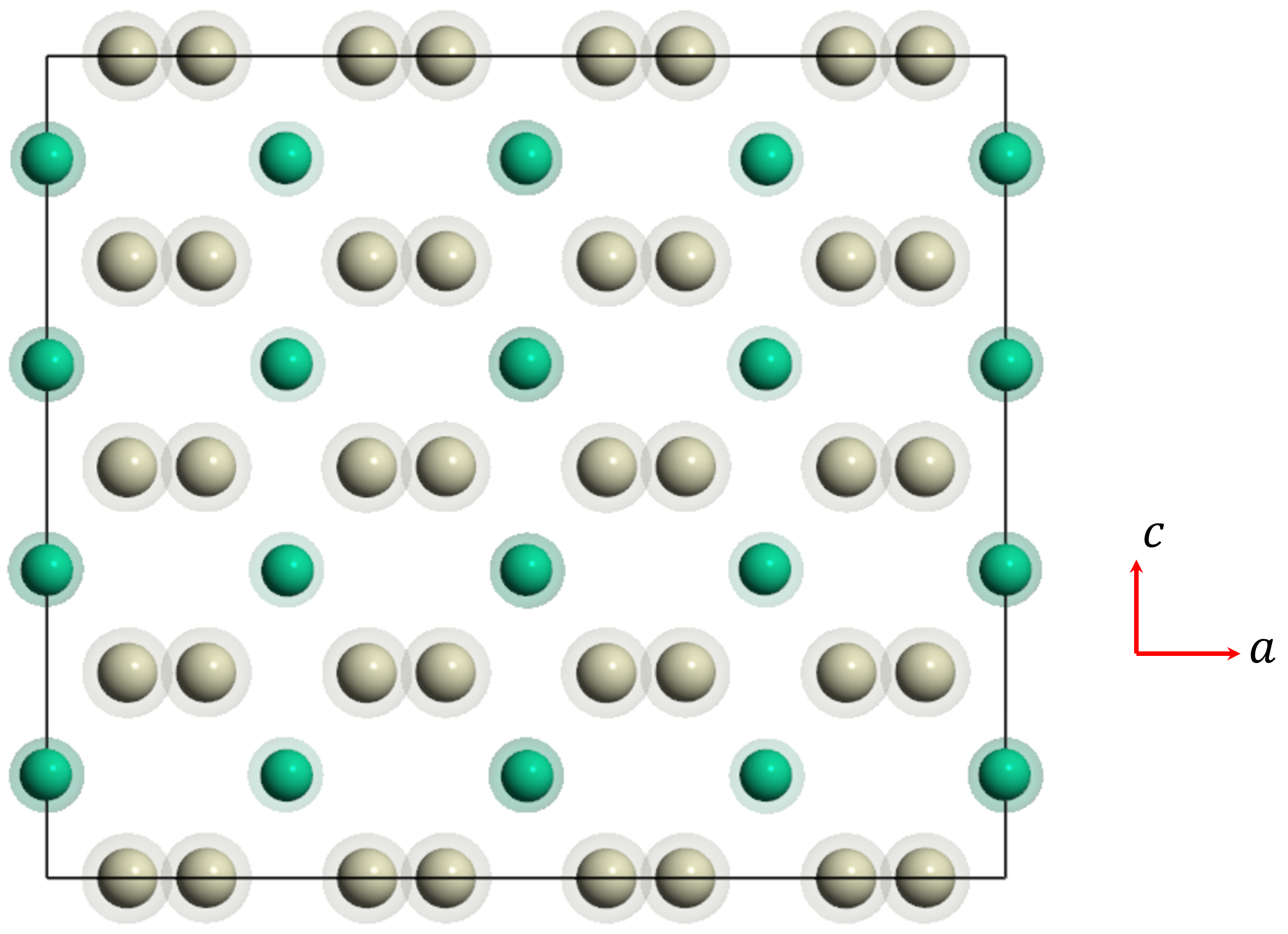}
		\caption{Inverted view of the 22~K structure overlapped with original structure at 22~K. Faint atoms represent the original atomic positions at 22~K before inversion. The Sn atoms preserve inversion symmetry and the Co atoms break inversion symmetry, as indicated by the slight displacement between their original and inverted positions.}
		\label{FigureS4}
	\end{figure*}
Figure \ref{FigureS4} provides a view for inversion symmetry breaking in CoSn$_2$ at 22~K. The figure overlays the original 22~K crystal structure with its inverted version, obtained by applying inversion through the center of the unit cell. Absence of center of inversion can be understood due to lack of superimposition particularly among the Co atoms (green spheres). This displacement indicates that Co atoms are primarily responsible for breaking inversion symmetry in the low temperature phase unlike  the Sn atoms which largely retain symmetric positions under inversion. 

\section{Magnetoresistance}
\begin{figure*}[!ht]
		\centering
		\includegraphics[scale=0.4]{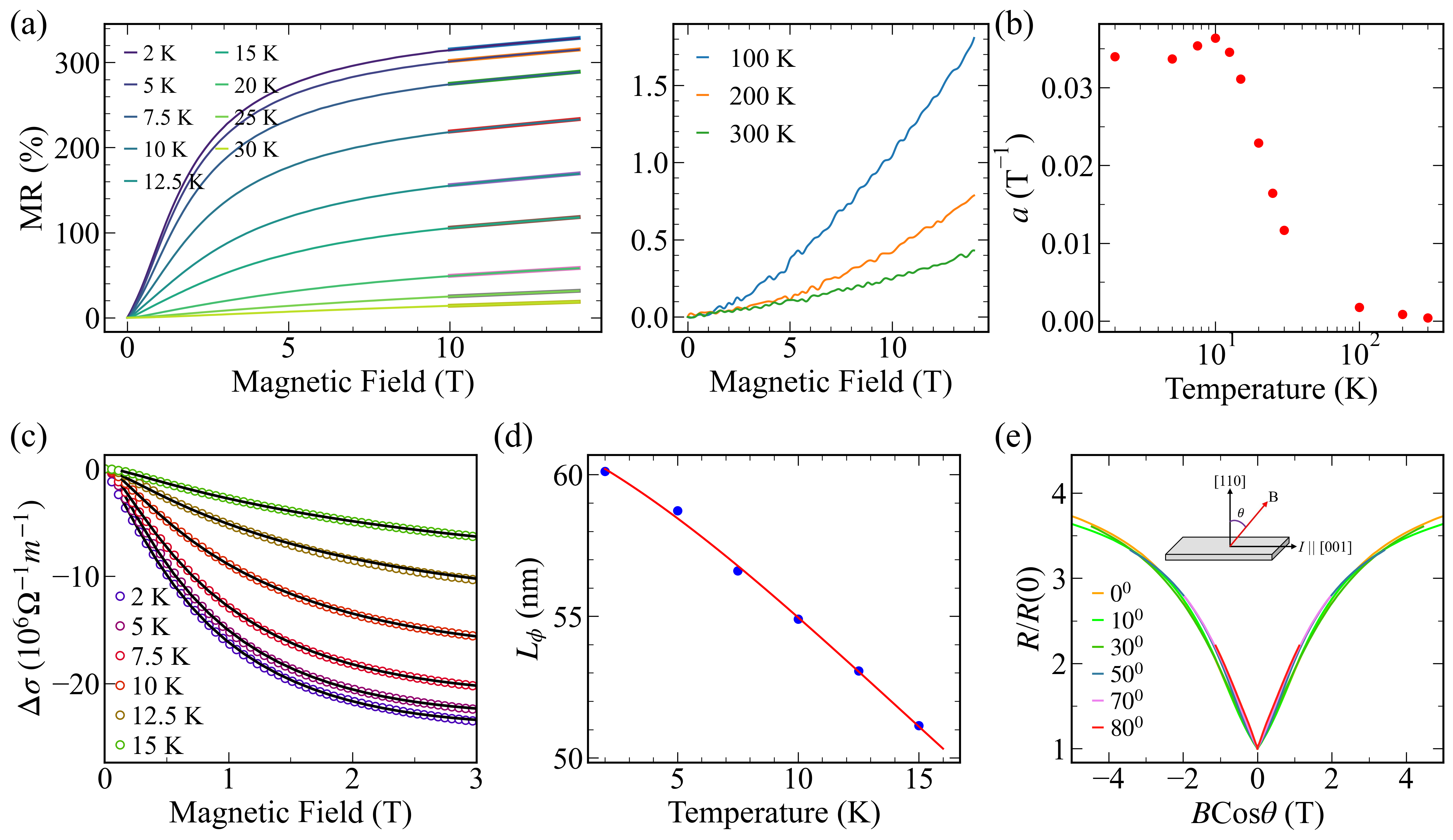}
		\caption{(a) Magnetoresistance (MR) measured at various temperatures. (b) Temperature dependence of the linear component of resistivity. (c) Magnetoconductivity at different temperatures; solid lines represent fits to Eq.~\ref{eqn5.4}. (d) Phase coherence length ($L_\phi$) as a function of temperature. (e) Normalized resistance plotted against $B\cos\theta$ at different tilt angles between the current and magnetic field.}
		\label{Figure5.6}
	\end{figure*}
Figure~\ref{Figure5.6}(a) shows the magnetic field dependence of magnetoresistance with the current along the [001] direction and the magnetic field parallel to the [110] direction at selected temperatures. A pronounced cusp-like behavior is observed in the low-field region of the MR below 20 K, which resembles the characteristics of weak antilocalization (WAL). WAL occurs due to destructive interference between two time-reversed electron paths. This effect is commonly observed in systems with strong spin-orbit coupling (SOC) or in those featuring two-dimensional (2D) surface states. In order to describe WAL in a 3D system, we adopt a generalized approach, analogous to the Hikami, Larkin, and Nagaoka model for 2D systems, as given by~\cite{Nakamura2020}:
	
\begin{equation}
		\Delta\sigma(B)=\frac{N}{4\pi h l_B}\left[2\xi\left(\frac 1 2,\frac 1 2 +\frac{l_B^2}{l^2}\right)+\xi\left(\frac12,\frac 1 2 +\frac{l_B^2}{l_\phi^2}\right)-3\xi\left(\frac 1 2,\frac 1 2+4\frac{l_B^2}{l_{SO}^2} +\frac{l_B^2}{l_\phi^2}\right)\right]
		\label{eqn5.3}
\end{equation}
	Here, N is the number of independent interference channels, $\xi$ is
	the Hurwitz zeta function. The length
	scales $l_B=\sqrt{\hbar/4eB}$ , $l$, $l_\phi$, and $l_{SO}$ denote the magnetic length, the mean free path, the phase coherence length, and the spin–orbit scattering length, respectively. As hall resistivity is negligible with respect to linear resistivity, we have calculated $\Delta\sigma$ as $1/\rho(B)-1/\rho(0)$ and plotted with field in Fig.~\ref{Figure5.6}(c). Eq.~\label{eqn5.3} shows a good fit to the experimental $\Delta\sigma$ at low fields. At $T$=2 K, the fitting yields $l$=12 nm, $l_\phi$=60 nm, and $l_{SO}$=34 nm. The key criterion for observing WAL, $l_\phi > l$, is satisfied in this case, and the spin-orbit coupling is not too strong. As depicted in Fig.~\ref{Figure5.6}(d), the $l_\phi$ data does not seem to follow the expected $T^{-0.5}$ or $T^{-0.75}$ dependence, suggesting the involvement of multiple scattering mechanisms for dephasing of the electron phase. We attempt to analyze the temperature dependence of $l_\phi$ using the following simplified equation:
	\begin{equation}
		\frac{1}{l_\phi^2}=\frac{1}{l_0^2}+A_{ee}T+A_{ep}T^2
		\label{eqn5.4}
	\end{equation}
	
	where $l_0$ is the zero temperature dephasing length, and $A_{ee}T$ and $A_{ep}T^2$ are the electron-electron and electron-phonon contribution, respectively. The extracted fitting parameters are $l_0 = 53.7$ nm, $A_{ee} = 3.35 \times 10^{-7}{\rm nm^{-2}K^{-1}}$, and $A_{ep} = 2.32 \times 10^{-7}{\rm nm^{-2}K^{-2}}$, indicating that electron-phonon interaction is the dominant dephasing mechanism. To clarify whether the WAL effect arises from 2D topological surface states or bulk states, we investigated the angular dependence of normalized resistance $R(B,\theta)/R(0,\theta)$ using the rotational geometry depicted in Fig.~\ref{Figure5.6}(e). We plotted the normalized resistance as a function of the perpendicular field component, $B\cos{\theta}$ (Fig.~\ref{Figure5.6}(e)), and observed that the data at various $\theta$ angles do not exactly collapse onto a single curve. This deviation indicates a significant contribution from bulk states, possibly due to moderate SOC. Apart from the low field WAL behaviour we observe a large linear magnetoresintance (LMR) at high fields and low temperatures. However, at higher temperatures, the magnetoresistance exhibits a quadratic behavior instead of linear, as shown in right panel of Fig.~\ref{Figure5.6}(a). The linear component of the resistivity changes sharply above 10~K (Fig.~\ref{Figure5.6}(b)). Two primary physical models have been proposed to explain the linear magnetoresistance (LMR) observed in various systems. The classical Parish and Littlewood (PL) model~\cite{PhysRevB.72.094417} attributes the linearity in magnetoresistance to inhomogeneity or disorder within the system. 
	However, this explanation is unlikely in our case, given the high crystalline quality of our single crystals, which rules out significant disorder. In contrast, Abrikosov introduced a quantum model for LMR~\cite{PhysRevB.58.2788}, predicting that in the extreme quantum limit, charge carriers occupy the lowest degenerate Landau level. Yet, the absence of any observable quantum oscillations in our data argues against this mechanism as well. An alternative explanation involves the presence of Dirac fermions, which can lead to a linear magnetoresistance due to their unique electronic structure and linear dispersion~\cite{PhysRevB.102.235431,Tang2011}. In our ARPES measurements, we observe linearly dispersing Dirac-like states near the Fermi level, consistent with this scenario.

\section{Photon energy dependent ARPES}
\begin{figure*}[!ht]
		\centering
		\includegraphics[scale=0.48]{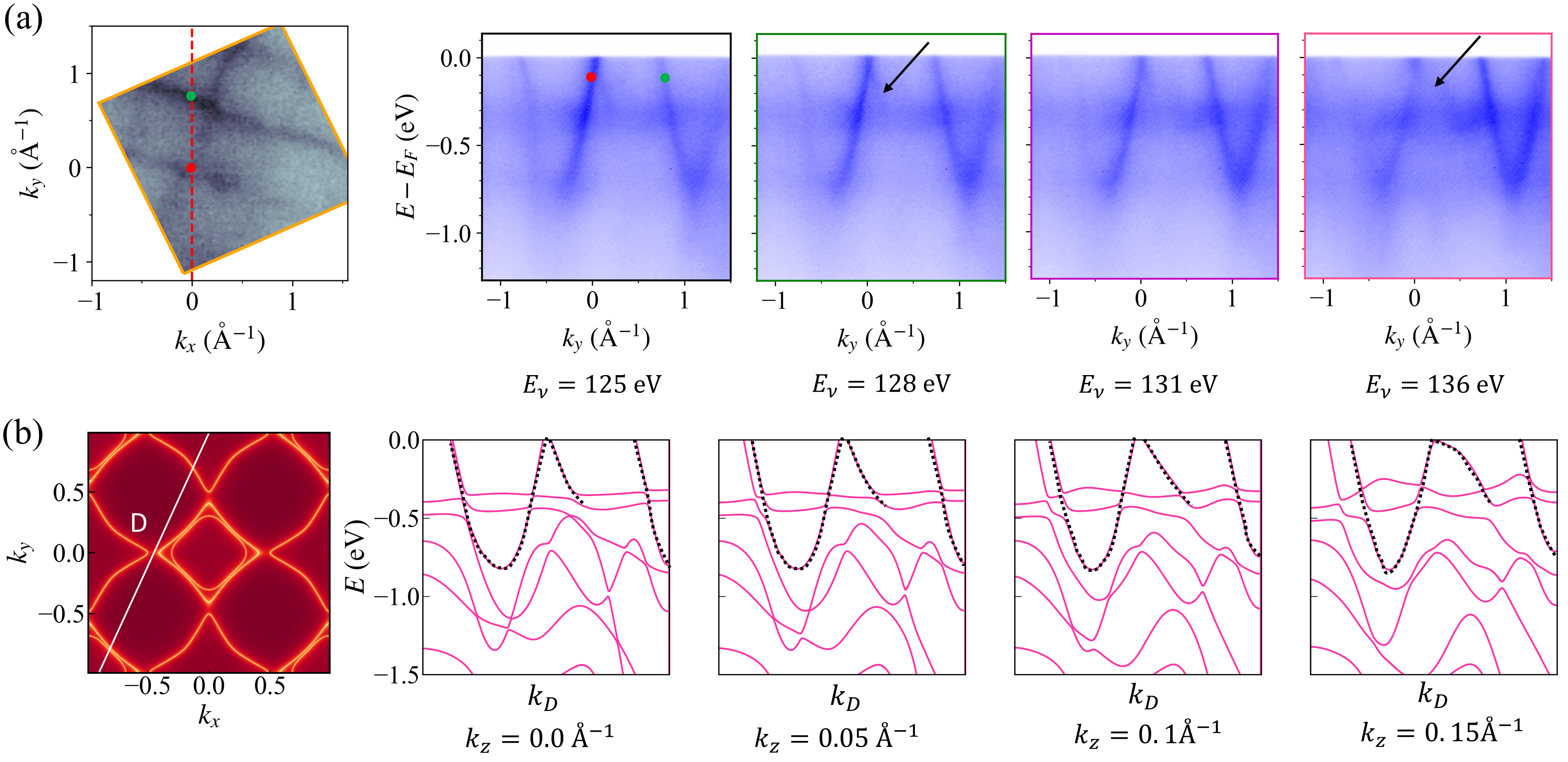}
		\caption{(a) Band dispersion along red line at different photon energies. Arrows indicate the evolution of Dirac crossing. (b) Left: Calculated Fermi surface highlighting the measured cut direction (white line). Right: band dispersions along $k_D$ for different $k_z$ values. The dotted black lines serve as guides to the experimental dispersions.}
		\label{FigureS5}
	\end{figure*}
To check the band dispersion along different $k_z$
values, we varied the photon energy in the ARPES measurements. In this case, the sample was not oriented along a high-symmetry direction as shown in Fig.~\ref{FigureS5}(a). The four ARPES spectra to the right show dispersions along the $k_y$ direction at different photon energies (125,128,131,136 eV), which correspond to different $k_z$ values due to the photon energy dependence of the photoelectron’s out of plane momentum. However, the Dirac-like dispersion remains clearly visible across all measurements. Due to the high photon energy range used, a relatively large change in photon energy is required to cover a significant portion of the $k_z$ momentum space. To compare these experimental results with theory, we calculated the band dispersion using DFT along the $k_D$ path (Fig.~\ref{FigureS5}(b)), which is chosen to be close to the red dashed path shown in Fig.~\ref{FigureS5}(a). Despite the off symmetry orientation, the experimental data show good agreement with the DFT calculated bands, particularly the evolution of the Dirac-like feature with increasing $k_z$.

\end{document}